\documentclass[twocolumn,tradiabstract]{aa}
\usepackage{graphicx}
\usepackage{amsbsy}
\usepackage{natbib}
\usepackage{color}
\usepackage{txfonts}
\usepackage{url}
\usepackage{bm}

\newcommand{\de}{\mathrm{d}}

\DeclareMathSymbol{\varOmega}{\mathord}{letters}{"0A}
\DeclareMathSymbol{\varSigma}{\mathord}{letters}{"06}
\DeclareMathSymbol{\varPsi}{\mathord}{letters}{"09}
\DeclareMathSymbol{\varPhi}{\mathord}{letters}{"08}
\DeclareMathSymbol{\varGamma}{\mathord}{letters}{"00}

\begin{document}

\title{How planetary growth outperforms migration}
\titlerunning{How planetary growth outperforms migration}

\author{Anders Johansen\inst{1}, Shigeru Ida\inst{2}, and Ramon Brasser\inst{2}}
\authorrunning{Johansen, Ida, Brasser}

\offprints{\\A.\ Johansen (\email{anders@astro.lu.se})}

\institute{$^1$ Lund Observatory, Department of Astronomy and Theoretical
Physics, Lund University, Box 43, 221 00 Lund, Sweden, \\e-mail:
\url{anders@astro.lu.se} \\ $^2$ Earth-Life Science Institute (ELSI), Tokyo
Institute of Technology, Meguro, Tokyo, 152-8550, Japan}

\date{}

\abstract{Planetary migration is a major challenge for planet formation
theories. The speed of Type I migration is proportional to the mass of a
protoplanet, while the final decade of growth of a pebble-accreting planetary
core takes place at a rate that scales with the mass to the two-thirds power.
This results in planetary growth tracks (i.e., the evolution of a protoplanet's
mass versus its distance from the star) that become increasingly horizontal
(migration-dominated) with rising mass of the protoplanet. It has been shown
recently that the migration torque on a protoplanet is reduced proportional to
the relative height of the gas gap carved by the growing planet. Here we show
from 1-D simulations of planet-disc interaction that the mass at which a planet
carves a 50\% gap is approximately 2.3 times the pebble isolation mass. Our
measurements of the pebble isolation mass from 1-D simulations match published
3-D results relatively well, except at very low viscosities ($\alpha<10^{-3}$)
where the 3-D pebble isolation mass is significantly higher, possibly due to
gap edge instabilities that are not captured in 1-D. The pebble isolation
mass demarks the transition from pebble accretion to gas accretion. Gas
accretion to form gas-giant planets therefore takes place over a few
astronomical units of migration after reaching first the pebble isolation mass
and, shortly after, the 50\% gap mass. Our results demonstrate how planetary
growth can outperform migration, both during core accretion and during gas
accretion, even when the Stokes number of the pebbles is small, ${\rm
St}\sim0.01$, and the pebble-to-gas flux ratio in the protoplanetary disc is in
the nominal range of $0.01$--$0.02$.  We find that planetary growth is very
rapid in the first million years of the protoplanetary disc and that the
probability for forming gas-giant planets increases with the initial size of the
protoplanetary disc and with decreasing turbulent diffusion.}


\maketitle

\section{Introduction}

The formation of planets involves many distinct steps in the growth from dust
and ice particles to full planetary system \citep[a realisation that was
pioneered 50 years ago by][]{Safronov1969}. The competition between radial
migration and growth is nevertheless a general theme in planet formation.  The
formation of planetesimals, for example, has to overcome the radial drift caused
by the head-wind of the slower moving gas on the growing particles
\citep{Weidenschilling1977,Brauer+etal2008}. Trapping pebbles in pressure bumps
formed in the turbulent gas flow
\citep{Lyra+etal2008,Johansen+etal2009a,Bai2015} or by embedded planets
\citep{Lyra+etal2009} or through run-away pile-ups caused by the friction on the
gas
\citep{YoudinGoodman2005,Johansen+etal2009b,BaiStone2010,Gonzalez+etal2017,Drazkowska+etal2016}
provide possible solutions to the radial drift problem of planetesimal
formation.

Planetary migration results from the gravitational torque exerted on a
protoplanet by the spiral density wakes excited in the gaseous protoplanetary
disc \citep{LinPapaloizou1979,GoldreichTremaine1980,Tanaka+etal2002}. The
nominal Type I migration rate reaches approximately 1 m/s (about 100 AU/Myr) for
a 10-Earth-mass protoplanet embedded in a young, massive protoplanetary disc.
The solutions proposed to mitigate catastrophic Type I migration include slowing
down or even reversing the migration where the temperature gradient is steep in
the inner regions of the protoplanetary disc
\citep{Paardekooper+etal2010,Lyra+etal2010,Bitsch+etal2014,Brasser+etal2017}, or
by a positive corotation torque resulting from the accretion heat of the
protoplanet \citep{Benitez-Llambay+etal2015} or from the scattering of (large)
pebbles \citep{Benitez-LlambayPessah2018}. Migration can even stall entirely if
the turbulent viscosity is too weak to diffuse away the gas density enhancement
that forms interior of the planetary orbit
\citep{Rafikov2002,Li+etal2009,FungLee2018}.

A more direct way to overcome Type I migration is simply for the planetary core
to increase its mass very rapidly. The accretion rate of pebbles is potentially
high enough for the protoplanet growth to outperform the nominal rate of Type I
migration \citep{LambrechtsJohansen2014,Ormel+etal2017,JohansenLambrechts2017}.
\cite{Bitsch+etal2015b} demonstrated that protoplanets forming several tens of
AU from the host star have enough space to grow to gas-giant planets while they
migrate into 5--10 AU orbits\footnote{\cite{Matsumura+etal2017}, on the other
hand, started the seeds in the 5--10 AU region and showed how this leads mainly
to the formation of warm and hot Jupiters}. However, \cite{Brugger+etal2018}
found that the pebble flux rate in the model of \cite{Bitsch+etal2015b} was
artificially high and that nominal pebble fluxes ($\sim$$0.01$ relative to the
gas flux) do not yield high enough core growth rates to compete with migration,
unless the metallicity is enhanced significantly beyond the solar value.
\cite{Bitsch+etal2018a} argued instead that the pebble sizes and surface
densities of \cite{Bitsch+etal2015b} in fact correspond well to observations of
protoplanetary discs. However, although the protoplanetary disc model advocated
in \cite{Bitsch+etal2018a} is motivated observationally, the pebble sizes and
surface densities used in that work are not anchored in self-consistent
theoretical models of pebble growth and drift.

In this paper we develop pebble accretion models with small pebbles of Stokes
number ${\rm St}\sim0.01$ (approximately mm-sized in the planet formation
region), much smaller than those considered in \cite{Bitsch+etal2015b} and
\cite{Brugger+etal2018}. Such small pebbles follow the viscous accretion flow of
the gas and, in contrast to larger pebbles, their column density is not notably
reduced by radial drift \citep{LambrechtsJohansen2014}. We demonstrate how in
this model planetary formation outperforms migration for nominal pebble fluxes
and metallicities. We derive analytical expressions that describe the growth
tracks of solid protoplanets undergoing pebble accretion (Section 2). We find
that planetary cores do undergo substantial migration before reaching the pebble
isolation mass. However, using novel prescriptions for the migration rate of
gap-opening planets \citep{Kanagawa+etal2018}, we show that gas accretion to
form gas-giant planets takes place over just a few astronomical units of
migration (Section 3).  Hence the main migration phase of a protoplanet happens
during the accumulation of the core. We summarise our results and discuss the
implications for planet formation in Section 4. In Appendix A we discuss
the survival of pebbles in protoplanetary discs and in Appendix B we derive
approximate times associated with the analytical growth tracks from Section 2.
Our paper forms a companion paper to \cite{Ida+etal2018}, which focuses on the
effect of the \cite{Kanagawa+etal2018} migration rate on the gas accretion
stage.

\section{Analytical pebble accretion growth tracks}
\label{s:core_growth}

In this section we derive analytical expressions for the growth track of a
protoplanet growing by pebble accretion. We then use this expression to derive
the location where the growth track crosses the pebble isolation mass.

\subsection{Pebble accretion}

To derive the analytical shape of the growth track of a solid protoplanet, we
use the protoplanet growth rate from pebble accretion in the 2-D Hill regime
\citep{OrmelKlahr2010,LambrechtsJohansen2012,Ida+etal2016},
\begin{equation}
  \dot{M} = 2 \left( \frac{\rm St}{0.1} \right)^{2/3} \varOmega R_{\rm H}^2
  \varSigma_{\rm p} \, .
  \label{eq:Mdot}
\end{equation}
Here $M$ is the mass of the protoplanet, ${\rm St}=\varOmega \tau_{\rm f}$ is
the Stokes number of the pebbles (defined later in equation \ref{eq:St}),
$\varOmega$ is the Keplerian frequency at the location of the protoplanet,
$\tau_{\rm f}$ is the friction time of the pebble (proportional to the particle
size when pebbles are small), $R_{\rm H}=[M/(3 M_\star)]^{1/3} r$ is the Hill
radius of the protoplanet, $M_\star$ is the mass of the central star, $r$ is the
radial location of the planet and $\varSigma_{\rm p}$ is the pebble surface
density. Significant effort is currently being put into understanding how pebble
accretion depends on the eccentricity and inclination of the protoplanet
\citep{Johansen+etal2015,LiuOrmel2018} as well as on realistic hydrodynamical
flow in the Hill sphere \citep{Popovas+etal2018}, but we consider here the
relatively simpler case of a protoplanet on a circular orbit and the gas flow as
pure shear.

Our assumption of 2-D Hill accretion is valid when the pebble accretion radius,
$R_{\rm acc}=({\rm St}/0.1)^{1/3} R_{\rm H}$, is larger than the scale-height of
the pebble layer, $H_{\rm p}=H \sqrt{\delta/{\rm St}}$ \citep[see][for a
discussion]{Morbidelli+etal2015}. Here $H$ is the scale-height of the gas and
$\delta$ is the dimensionless dust diffusion coefficient \citep[defined
in][]{Johansen+etal2014}.  That gives an accretion radius relative to the pebble
scale-height as
\begin{equation}
  \frac{R_{\rm acc}}{H_{\rm p}} = 0.9 \left( \frac{\rm St}{0.01} \right)^{1/3}
  \left( \frac{M}{M_{\rm E}} \right)^{1/3} \left( \frac{H/r}{0.05}
  \right)^{-1} \left( \frac{{\rm St}/\delta}{100} \right)^{1/2} \, .
  \label{eq:Racc}
\end{equation}
We normalised here to ${\rm St}=0.01$ and $\delta=0.0001$. Our choice of Stokes
number ${\rm St}=0.01$ is elaborated in Section \ref{s:pebbles}, while the
choice of a low diffusion coefficient $\delta$ is motivated by observations of
dusty protoplanetary discs that show that the dust has settled to a mid-plane
layer of width 10\% of a gas scale-height
\citep{MuldersDominik2012,Menu+etal2014,Pinte+etal2016}. Note that ${\rm
St}/\delta=100$ indeed yields a moderately sedimented pebble mid-plane layer
with a scale-height relative to the gas scale-height of $H_{\rm p}/H=0.1$.

The transition from the Bondi regime (where the pebble approach speed is set by
the sub-Keplerian motion) to the Hill regime (where the pebble approach speed is
set by the Keplerian shear) of pebble accretion typically happens at 0.001-0.01
Earth masses \citep{LambrechtsJohansen2012}. Protoplanets that accrete either in
the Bondi regime or in the 3-D Hill regime thus experience very low migration
rates and hence the shape of the growth track is not affected by our choice to
start the analytical core growth tracks in the 2-D Hill accretion phase. In
Section \ref{s:full_growth} we include the 3-D branch of Hill accretion in the
numerical integrations.

\cite{JohansenLambrechts2017} demonstrated that the accretion of planetesimals
contributes significantly to the growth from planetesimal sizes to protoplanets
with masses $0.001-0.01\,M_{\rm E}$, due to the low rates of pebble accretion in
the Bondi regime, while pebble accretion in the Hill regime dominates the
further growth. Therefore we start our protoplanets at $M_0=0.01\,M_{\rm E}$ and
ignore the contribution of planetesimal accretion to the core growth rate,
although the inclusion of planetesimal accretion could give a boost to the
accretion rate, depending on the migration speed of the protoplanet
\citep{TanakaIda1999}.

\subsection{Type I migration}

For Type I migration we use the standard scaling law,
\begin{equation}
  \dot{r} = -k_{\rm mig} \frac{M}{M_\star} \frac{\varSigma_{\rm g}
  r^2}{M_\star} \left( \frac{H}{r} \right)^{-2} v_{\rm K} \, .
  \label{eq:rdot}
\end{equation}
Here $\dot{r}$ is the migration speed of the protoplanet, $k_{\rm mig}$ is a
constant prefactor that depends on the gradients of surface density and
temperature, $M_\star$ is the mass of the central star, $\varSigma_{\rm g}$ is
the gas surface density, $H/r$ is the disc aspect ratio and $v_{\rm K}$ is the
Keplerian speed at the position of the planet. For the prefactor $k_{\rm mig}$
we follow here a fit to 3-D numerical simulations found in
\cite{DAngeloLubow2010},
\begin{equation}
  k_{\rm mig} = 2 (1.36 + 0.62 \beta + 0.43 \zeta) \, ,
\end{equation}
where $\beta$ and $\zeta$ are the negative logarithmic derivatives of the
surface density and the temperature profiles, respectively. However, we
explicitly maintain $k_{\rm mig}$ in our equations, since the prefactor depends
on the physical effects that are included in the simulations
\citep[e.g.][]{Tanaka+etal2002,Paardekooper+etal2010}.

\subsection{Radial mass fluxes of gas and pebbles}

The gas sound speed $c_{\rm s}$ and the derived gas scale-height $H=c_{\rm
s}/\varOmega$ enter both the calculation of the planetary migration migration
rate as well as expressions for the radial speed of the gas and pebbles that we
present below. We assume that the sound speed profile follows a power law
\begin{equation}
  c_{\rm s} = c_{\rm s1} \left( \frac{r}{\rm AU} \right)^{-\zeta/2} \, .
\end{equation}
Here $\zeta$ is the negative power-law index of the temperature (proportional to
$c_{\rm s}^2$) and $c_{\rm s1}$ is the sound speed at 1 AU. The disc aspect
ratio then follows the power-law
\begin{equation}
  \frac{H}{r} \propto r^{-\zeta/2+1/2} \, .
  \label{eq:Hr}
\end{equation}

The turbulent viscosity $\nu$ sets the radial gas accretion speed. We use here
the $\alpha$-disc assumption for the turbulent viscosity
\citep[e.g.][]{Pringle1981},
\begin{equation}
  \nu = \alpha c_{\rm s} H \, .
  \label{eq:nu}
\end{equation}
This now results in the gas accretion speed
\begin{equation}
  u_r = -\frac{3}{2} \frac{\nu}{r} = -\frac{3}{2} \alpha c_{\rm s} \frac{H}{r}
  \, .
  \label{eq:ur}
\end{equation}
This expression for the accretion speed is specific to the $\alpha$-disc
assumption and would not be valid if the angular momentum loss was transported
instead by disc winds \citep{BaiStone2013}. However, since the radial mass
accretion rate of the gas depends only on the speed of the gas, and not on the
nature of the angular momentum transport, we can consider $\alpha$ in equation
(\ref{eq:ur}) simply a dimensionless measure of the radial accretion speed.

The radial drift of the particles is given by \citep{Weidenschilling1977}
\begin{equation}
  v_r = - \frac{2 \Delta v}{{\rm St}+{\rm St}^{-1}} + \frac{u_r}{1+{\rm St}^2}
  \, .
\end{equation}
In the limit ${\rm St} \ll 1$, valid for pebbles smaller than approximately
0.1-1 meters in size outside of a few AU \citep{Johansen+etal2014}, this
expression simplifies to
\begin{equation}
  v_r = - 2 {\rm St} \Delta v + u_r
\end{equation}
Here the sub-Keplerian speed $\Delta v$, which is a measure of the radial
pressure support of the gas, is given by
\begin{equation}
  \Delta v = -\frac{1}{2} \frac{H}{r} (\partial \ln P/\partial
  \ln r) c_{\rm s} \, .
\end{equation}
We denote the negative logarithmic pressure gradient in the mid-plane as
$-\partial \ln P/\partial \ln r \equiv \chi = \beta + \zeta/2 + 3/2$. The
inwards mass fluxes of gas and pebbles, respectively, are given by
\begin{eqnarray}
  \dot{M}_{\rm g} &=& -2 \pi r u_r \varSigma_{\rm g} \, ,  \\
  \dot{M}_{\rm p} &=& -2 \pi r v_r \varSigma_{\rm p} \, . \label{eq:Mdotp}
\end{eqnarray}
The ratio of the surface densities of pebbles to gas is then
\begin{equation}
  \frac{\varSigma_{\rm p}}{\varSigma_{\rm g}} = \frac{\dot{M}_{\rm
  p}}{\dot{M}_{\rm g}} \frac{u_r}{-2 {\rm St} \Delta v+u_r}
  = \frac{\xi}{(2/3)({\rm St}/\alpha) \chi + 1} \, .
  \label{eq:Zp}
\end{equation}
Here we defined the ratio of the fluxes as in \cite{Ida+etal2016},
\begin{equation}
  \xi = \frac{\dot{M}_{\rm p}}{\dot{M}_{\rm g}} \, .
\end{equation}
We will show below in Sections \ref{s:anagrowth} and \ref{s:isomass} that $\xi$
is a key parameter that determines the shape of a protoplanet's growth track and
the total migration distance of the protoplanet before reaching the pebble
isolation mass. The gas and solid components of the protoplanetary disc will
accrete towards the star at the same time-scale when $\xi \approx 0.01$, where
$0.01$ represents the metallicity $Z$ of the protoplanetary disc. This nominal
value of $\xi$ is obtained when the radial drift of the pebbles is dominated by
advection with the accreting gas, corresponding to ${\rm St}/\alpha \lesssim 1$
in equation (\ref{eq:Zp}). Therefore also the local metallicity $\varSigma_{\rm
p}/\varSigma_{\rm g}$ will keep its original value $Z$ for the nominal
pebbles-to-gas flux ratio.

Large pebbles with ${\rm St} \gg \alpha$ in equation (\ref{eq:Zp}) experience an
increase in $\xi$ proportional to ${\rm St}$ and hence the local metallicity
$\varSigma_{\rm p}/\varSigma_{\rm g}$ is maintained at its global value $Z$. An
increased Stokes number will nevertheless have an overall positive effect on the
pebble accretion rate (through equation \ref{eq:Mdot}), but such large pebbles
are lost to radial drift on a shorter time-scale than the gas accretion. This is
the well-known radial drift problem of protoplanetary discs
\citep{Brauer+etal2007}. Another possibility is that $\xi$ is dictated by the
production rate of pebbles in the outer regions of the protoplanetary disc
\citep{LambrechtsJohansen2014,Bitsch+etal2015b}. In that case $\xi$ is no longer
directly coupled to ${\rm St}$ in equation (\ref{eq:Zp}) and the local
metallicity $\varSigma_{\rm p}/\varSigma_{\rm g}$ falls proportional to the
inverse Stokes number when ${\rm St} \gg \alpha$. Overall, there are then many
advantages to considering small pebbles for pebble accretion models, in contrast
to the large pebbles that were used in the models of \cite{Bitsch+etal2015b} and
\cite{Brugger+etal2018}.

\subsection{Pebble sizes}
\label{s:pebbles}

Dust in protoplanetary discs grows to pebbles through coagulation
\citep{Brauer+etal2007,Zsom+etal2010}. If particles stick perfectly when they
collide, then growth continues until the radial drift time-scale becomes
comparable to the growth time-scale, at Stokes numbers around 0.1--1 in the
region of giant planet formation
\citep{Birnstiel+etal2012,LambrechtsJohansen2014}. Pebble accretion with such
drift-limited pebble growth was explored in \cite{Bitsch+etal2015a} and
\cite{Brugger+etal2018}.

Here we focus instead on a different, and perhaps more realistic, mode of pebble
growth where the pebbles experience bouncing or fragmenting collisions.
\cite{Zsom+etal2010} showed that the growth of silicate particles is limited to
millimeter sizes by bouncing, based on extensive experimental data on collisions
between silicate dust aggregates. Such experiments also show that collisions
become fragmenting when the collision speed crosses a threshold value
\citep{Birnstiel+etal2012}.

Water ice, in contrast to silicates, has higher surface energy and is thus
expected to experience growth beyond the bouncing barrier
\citep{Okuzumi+etal2012}, while CO$_2$ ice (and likely CO ice as well) appears
to have sticking properties similar to silicates
\citep{Musiolik+etal2016a,Musiolik+etal2016b}.  The CO$_2$ ice line sits at a
temperature of approximately 70 K at solar abundances, corresponding to the 2--4
AU region in the late stages of protoplanetary disc evolution
\citep{Bitsch+etal2015a,Madhusudhan+etal2017}. Hence, we expect that the growth
of pebbles is limited by bouncing or fragmentation in the accretion region of
the cores of cold gas giants. Ices of CO and CO$_2$ may, in turn, under UV
irradiation form longer refractory organic molecules
\citep{MunozCaroSchutte2003}. We ignore here the possibility that such organic
molecules could be sticky and facilitate the formation of pebbles larger than
millimeters in size \citep{Lodders2004}.

Both pebble accretion (in the Hill regime) and radial drift depend on the Stokes
number of the pebbles rather than on their physical sizes. The pebble Stokes
number is calculated from the radius $R$ and material density $\rho_\bullet$ of
the pebbles through the relation
\begin{equation}
  {\rm St} = \sqrt{2 \pi} \frac{R \rho_\bullet}{\varSigma_{\rm g}} = \sqrt{2
  \pi} \frac{3 \pi \nu R \rho_\bullet}{\dot{M}_{\rm g}} \, .
  \label{eq:St}
\end{equation}
This yields the pebble size for a given value of ${\rm St}/\alpha$ as
\begin{eqnarray}
  R &=& \frac{\rm St}{\alpha} \frac{\dot{M}_{\rm g}}{\sqrt{2\pi} 3 \pi c_{\rm s}
  H \rho_\bullet} \nonumber \\ &=& 1.1\,{\rm mm} \left( \frac{\rm St}{\alpha}
  \right) \left( \frac{\dot{M}_{\rm g}}{10^{-8}\,M_\odot\,{\rm yr}^{-1}} \right)
  \left( \frac{c_{\rm s1}}{6.5 \times 10^2\,{\rm m\,s^{-1}}} \right)^{-2}
  \nonumber \\
  && \times \left( \frac{\rho_\bullet}{10^3\,{\rm kg\,m^{-3}}} \right)^{-1}
  \left( \frac{r}{10\, {\rm AU}} \right)^{\zeta-3/2} \, .
\end{eqnarray}
In Section \ref{s:full_growth} we adopt a nominal protoplanetary disc evolution
model where the gas accretion rate drops from $10^{-7}\,M_\odot\,{\rm yr}^{-1}$
to $10^{-8}\,M_\odot\,{\rm yr}^{-1}$ over three million years
\citep{Hartmann+etal2016}. This gives a typical pebble size of mm-cm in the
5--20 AU region, for ${\rm St}/\alpha=1$, which corresponds well to the sizes of
pebbles whose growth is stuck at the bouncing barrier
\citep{Zsom+etal2010,Birnstiel+etal2012}.

The fragmentation barrier is reached when the collision speed driven by the
turbulent gas motion, $v_{\rm c} = \sqrt{3 \alpha_{\rm v} {\rm St}} c_{\rm s}$
\citep{OrmelCuzzi2007}, equals an assumed fragmentation speed, $v_{\rm f}$
\citep{Birnstiel+etal2012}. Here $\alpha_{\rm v}$ is the turbulent viscosity; we
discuss its value and connection to the global disc accretion coefficient
$\alpha$ in Section \ref{s:isomass}.  The turbulent collision speed results in a
limiting Stokes number
\begin{eqnarray}
  {\rm St} &=& \frac{1}{3} \alpha_{\rm v}^{-1} \left( \frac{v_{\rm f}}{c_{\rm
  s}} \right)^2 = 0.003 \left( \frac{\alpha_{\rm v}}{10^{-4}} \right)^{-1}
  \nonumber \\
  & & \times \left(
  \frac{v_{\rm f}}{1\,{\rm m\,s^{-1}}} \right)^2 \left( \frac{c_{\rm
  s1}}{6.5\times10^2\,{\rm m\,s^{-1}}} \right)^{-2} \left( \frac{r}{10\,{\rm
  AU}} \right)^{\zeta} \, .
\end{eqnarray}
The fragmentation barrier thus lies at a Stokes number that is (a) independent
of the temporally decaying mass accretion rate onto the star and (b) only weakly
dependent on the distance from the star.

The small pebbles stuck at the bouncing or fragmentation barrier have radial
speeds that are approximately the same as the radial gas accretion speed. This
results in a pebble-to-gas flux ratio $\xi \approx Z$, where $Z$ is the overall
metallicity of the protoplanetary disc, and hence a similar depletion time of
the gaseous and solid components of the protoplanetary disc. We will therefore,
for simplicity, consider ${\rm St}\sim\alpha$ in our models, as this choice
agrees both with the bouncing and fragmentation barriers and with observations
that show that pebbles of mm-cm sizes remain present in protoplanetary discs
over a wide range of ages \citep{Perez+etal2012,Huang+etal2018}. We discuss
the life-time of the pebble component of protoplanetary discs more in Appendix
\ref{s:pebble_disc}.

Choosing a larger value of ${\rm St}/\alpha$ would correspond to larger pebbles,
more in agreement with pebble growth limited only by the radial drift
\citep{Birnstiel+etal2012,LambrechtsJohansen2014}. Such large pebbles would lead
to high pebble accretion rates, but would drain out of the protoplanetary disc
on a shorter time-scale than the gas accretion time-scale \citep[][our Appendix
A]{Lin+etal2018}, adding significant complication in modelling the divergent
evolution of gas and pebbles.
\begin{figure*}
  \begin{center}
    \includegraphics[width=0.9\linewidth]{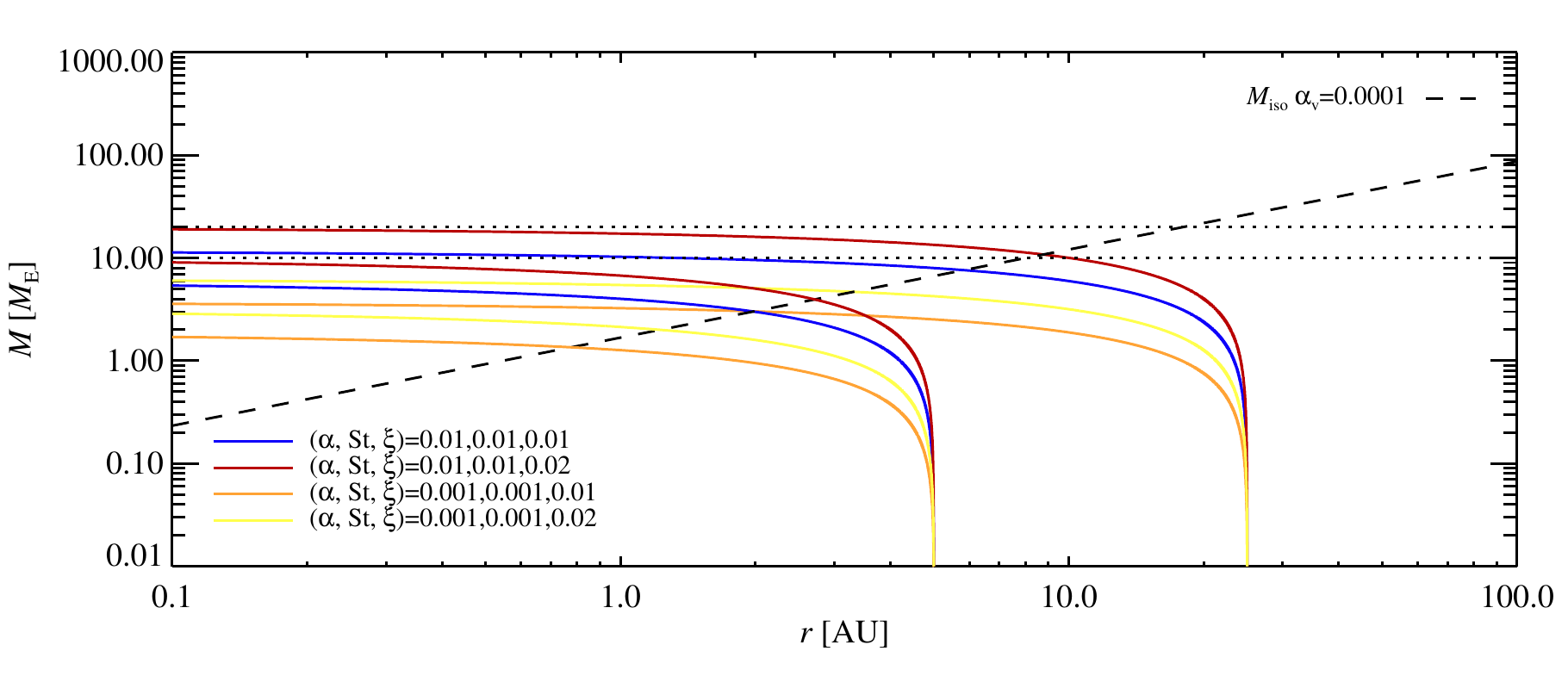}
  \end{center}
  \caption{Analytical growth tracks of planetary cores for four combinations of
  the accretion viscosity $\alpha$, the pebble Stokes number ${\rm St}$ (set to
  be equal to $\alpha$) and the ratio of radial pebble-to-gas flux rates $\xi$.
  Protoplanets are started at either 5 AU or 25 AU and with a starting mass of
  $0.01\,M_{\rm E}$. The pebble isolation mass is indicated for two values of
  the turbulent viscosity $\alpha_{\rm v}$, assumed here to be 0.1 times the
  accretion viscosity $\alpha$. The proportionality between the migration rate
  and the mass results in all growth tracks turning nearly horizontal before
  reaching their maximal mass at $r=0$.}
  \label{f:growth_tracks_analytical}
\end{figure*}

\subsection{Analytical core growth track}
\label{s:anagrowth}

Using equation (\ref{eq:Mdot}) for the core growth rate and equation
(\ref{eq:rdot}) for the migration rate, we can now formulate the differential
equation for the growth track $M(r)$,
\begin{eqnarray}
  \frac{{\rm d}M}{{\rm d}r} &=& \frac{\dot{M}}{\dot{r}} \nonumber \\
  &=& -\frac{\xi}{(2/3) ({\rm St}/\alpha) \chi + 1}
  \frac{2 ({\rm St}/0.1)^{2/3} M_\star (3 M_\star)^{-2/3}}{k_{\rm mig} G c_{\rm
  s1}^{-2} {\rm AU}^{-\zeta}} \nonumber \\
  && \times r^{-\zeta} M^{-1/3} \, .
  \label{eq:M43long}
\end{eqnarray}
The solution is found by separation of variables,
\begin{eqnarray}
  M^{4/3}-M_0^{4/3} &=& -\frac{(4/3) \xi}{(2/3)({\rm St}/\alpha) \chi + 1}
  \frac{2 ({\rm St}/0.1)^{2/3} M_\star (3 M_\star)^{-2/3}}{k_{\rm mig} G
  c_{\rm s1}^{-2} {\rm AU}^{-\zeta}} \nonumber \\
  && \times \frac{1}{1-\zeta}
  \left(r^{1-\zeta}-r_0^{1-\zeta}\right) \, .
\end{eqnarray}
Here $M_0$ and $r_0$ are the starting mass and starting location of the
protoplanet, respectively. We can now divide the equation by its solution at
$r=0$, $M(0)=M_{\rm max}$, to obtain
\begin{equation}
  \frac{M^{4/3}-M_0^{4/3}}{M_{\rm max}^{4/3}-M_0^{4/3}} = 1 - \left(
  \frac{r}{r_0} \right)^{1-\zeta} \, .
  \label{eq:M43}
\end{equation}
The ``maximum mass'' reached at $r=0$ for $\zeta<1$ is given by
\begin{eqnarray}
  M_{\rm max}^{4/3} &=& M_0^{4/3} + \frac{(4/3) \xi}{(2/3)({\rm St}/\alpha)\chi
  +1} \frac{2 ({\rm St}/0.1)^{2/3} M_\star (3
  M_\star)^{-2/3}}{k_{\rm mig} G c_{\rm s1}^{-2} {\rm AU}^{-\zeta}} \nonumber \\
  && \times \frac{r_0^{1-\zeta}}{1-\zeta} \, .
  \label{eq:Mmax}
\end{eqnarray}
For $\zeta>1$, the aspect ratio $H/r$ increases when approaching the star
(equation \ref{eq:Hr}) and migration is stalled by the high temperature in the
inner regions of the protoplanetary disc. The protoplanet therefore never
reaches $r=0$ in that case. We can reformulate equation (\ref{eq:Mmax}) as a
scaling law for $M_{\rm max}$ in the limit $M_{\rm max}\gg M_0$,
\begin{eqnarray}
  M_{\rm max} &=& 11.7\,M_{\rm E}\,\frac{({\rm
  St}/0.01)^{1/2}}{\{[(2/3)({\rm
  St}/\alpha) \chi +1]/2.9\}^{3/4}} \left(
  \frac{\xi}{0.01} \right)^{3/4} \nonumber \\ && \times \left(
  \frac{M_\star}{M_\odot} \right)^{1/4} \left(
  \frac{k_{\rm mig}}{4.42} \right)^{-3/4}
  \left(
  \frac{c_{\rm s1}}{6.5 \times 10^2 \,{\rm m\,s^{-1}}} \right)^{3/2} \nonumber
  \\ && \times \left( \frac{1-\zeta}{4/7} \right)^{-1} \left(
  \frac{r_0}{25\,{\rm AU}} \right)^{(3/4)(1-\zeta)} \, .
  \label{eq:Mmaxs}
\end{eqnarray}
We see here how $\xi$, which sets the pebble accretion rate, and $k_{\rm mig}$,
which sets the migration rate, pull the maximum mass in opposite directions at
an exactly equal power index.  Any reduction in the pebble accretion rate, e.g.\
from 3-D accretion if the pebble scale-height is larger than 10\% of the gas
scale-height in equation (\ref{eq:Racc}), will act in similar way as a reduction
in $\xi$ to reduce $M_{\rm max}$. We assumed in the derivations above that
$k_{\rm mig}$ is a constant.  This assumption may break down in the inner,
viscously heated regions of the protoplanetary disc (typically interior of 1 AU
for nominal accretion rates and disc masses) where the positive corotation
torque can slow down or reverse migration
\citep{BitschJohansen2016,Brasser+etal2017}. We note also that the weak scaling
with $M_\star^{1/4}$ in equation (\ref{eq:Mmaxs}) does not take into account
that $c_{\rm s1}$ depends on the luminosity, and hence on the mass, of the
star\footnote{Assuming that the luminosity of the host star has a power-law
dependence on the stellar mass, $L_\star \propto M_\star^p$, yields a
temperature at 1 AU that scales as $T_1 \propto L_\star^{2/7} M_\star^{-1/7}
\propto M_\star^{(2 p - 1)/7}$ \citep{Ida+etal2016} and hence $c_{\rm s1}
\propto M_\star^{(2 p-1)/14}$. That gives a combined mass-dependence of equation
(\ref{eq:Mmaxs}) as $M_{\rm max} \propto M_\star^{1/4 + (3 p - 3/2)/14}$, which
is close to linear for $p=3$.}

We finally obtain the shape of the growth track, $r(M)$, from equation
(\ref{eq:M43}),
\begin{equation}
  r(M) = r_0 \left( 1 - \frac{M^{4/3}-M_0^{4/3}}{M_{\rm max}^{4/3}-M_0^{4/3}}
  \right)^{1/(1-\zeta)} \, .
\end{equation}
In Appendix \ref{s:growth_time} we use this expression to derive the time associated with each
step in the growth track. We show analytical growth tracks in Figure
\ref{f:growth_tracks_analytical}, for pairs of values of $\alpha={\rm St}$ and
$\xi$. We use a disc temperature profile here with $c_{\rm s1}=650\,{\rm
m\,s^{-1}}$ and $\zeta=3/7$, appropriate for the outer regions of the
protoplanetary disc where viscous heating is negligible
\citep{ChiangYoudin2010,Bitsch+etal2015a,Ida+etal2016}. The growth tracks start
off nearly vertical (growth dominates over migration), but as the migration rate
increases, eventually the growth tracks turn horizontal and reach $M_{\rm max}$
after migrating to $r=0$. Type I migration is truly a tough opponent for the
planetary core to overcome.

\subsection{Pebble isolation mass}
\label{s:isomass}

The core growth stage ends as the planetary core reaches the pebble isolation
mass. Here the protoplanet's gravity perturbs the gas flow enough to form a
plateau of Keplerian motion on the exterior side of the orbit, trapping the
migrating pebbles there. The lack of heating by infalling pebbles then allows
the gas to decrease its entropy by radiative heat loss and contract slowly to
accrete a growing envelope around the core \citep{Lambrechts+etal2014}. In
\cite{Bitsch+etal2018b} the pebble isolation mass was fitted to 3-D simulations
by the expression
\begin{eqnarray}
  M_{\rm iso} &=& 25\,M_{\rm E} \left[ \frac{H/r}{0.05} \right]^3 \left[ 0.34
  \left( \frac{{\rm log}(\alpha_3)}{{\rm log}(\alpha_{\rm v})} \right)^4 + 0.66
  \right] \nonumber \\ && \times \left[ 1 - \frac{\partial \ln P/\partial \ln r
  + 2.5}{6} \right] \, .
  \label{eq:Miso}
\end{eqnarray}
Here $\alpha_3=10^{-3}$ is a constant and $\alpha_{\rm v}$ is the turbulent
viscosity, which we distinguish in this paper from the $\alpha$-value of the
mass accretion rate of equation (\ref{eq:nu}), as the latter may be driven by
disc winds at a weak level of actual turbulence
\citep{BaiStone2013,Bethune+etal2017}.  \cite{DurmannKley2015} showed that the
gas flow through the protoplanetary disc achieves a constant rate through a
planetary gap and that the migration rate is not dependent on the global gas
flow. We therefore here assume that the global gas speed also does not affect
gap formation and use a nominal value of $\alpha_{\rm v}=10^{-4}$ to
calculate the pebble isolation mass in equation (\ref{eq:Miso}) \citep[see ][for
a discussion of this distinction]{Armitage+etal2013,Hasegawa+etal2017}. We
motivate our usage of $\alpha_{\rm v}$ in defining the pebble isolation mass by
noting that the gas speed driven by the turbulent viscosity over the
length-scale of the gap, assumed to be the gas scale-height $H$, is
\begin{equation}
  u_{\rm v} \sim \frac{\nu}{H} \sim (H/r)^{-1} u_r \, ,
\end{equation}
where $u_r$ is the global gas accretion speed. Hence turbulent viscosity is
expected to be a factor $(H/r)^{-1}$ times more effective than the global gas
accretion speed at counteracting gap formation. However, we emphasize that
\cite{DurmannKley2015} did not distinguish between the $\alpha$ measuring disc
accretion and the $\alpha_{\rm v}$ measuring turbulent viscosity. Future studies
are needed to assess the effect of the global gas accretion speed, here
parameterised through $\alpha$, on gap formation and pebble isolation mass.

The isolation mass for the nominal value of $\alpha_{\rm v}$ is overplotted in
Figure \ref{f:growth_tracks_analytical}. The pebble isolation mass is increased
over the expression given in equation (\ref{eq:Miso}) for small pebbles or
strong particle turbulence \citep{Bitsch+etal2018b,Ataiee+etal2018}; we ignore
such effects here since we work in the limit of weak turbulence and since even
small pebbles of ${\rm St}=0.01$ are easily trapped at the outer edge of the
gap.

\subsection{Reaching the pebble isolation mass}

One can derive analytically the location where the protoplanet growth track
crosses the pebble isolation mass. The pebble isolation mass changes as $r$
decreases from its initial $r_0$. Scaling equation (\ref{eq:Miso}) from the
starting position of the growth track, $r_0$, we obtain the expression
\begin{equation}
  M_{\rm iso}(r) = M_{\rm iso,0} \left( \frac{r}{r_0} \right)^{(3/2)(1-\zeta)}
  \, .
\end{equation}
Here $M_{\rm iso,0}$ is the isolation mass at $r=r_0$. In the limit $M \gg M_0$
the protoplanet therefore reaches pebble isolation mass at $r=r_{\rm iso}$ when
\begin{equation}
  M_{\rm iso}^{4/3} = M_{\rm iso,0}^{4/3} \left( \frac{r_{\rm iso}}{r_0}
  \right)^{2 (1-\zeta)} = M_{\rm max}^{4/3} \left[ 1 - \left(
  \frac{r_{\rm iso}}{r_0} \right)^{1-\zeta} \right] \, .
  \label{eq:Miso0}
\end{equation}
This equation forms a second order polynomial,
\begin{equation}
  \left( \frac{M_{\rm iso,0}}{M_{\rm max}} \right)^{4/3} X^2 + X - 1 = 0 \, ,
  \label{eq:pol}
\end{equation}
with $X=(r_{\rm iso}/r_0)^{1-\zeta}$ or $\Delta r_{\rm iso} = r_0 - r_{\rm iso}
= [1-X^{1/(1-\zeta)}] r_0$. The positive solution to the polynomial is
\begin{equation}
  X=\frac{\sqrt{1+4A}-1}{2 A} \, ,
  \label{eq:X}
\end{equation}
where $A=(M_{\rm iso,0}/M_{\rm max})^{4/3}$. Note that a solution exists for all
$A$, i.e.\ the pebble isolation mass is reached for all starting positions
$r_0$. That is due to the steeply falling pebble isolation mass with decreasing
distance, in the passively irradiated case. Our results generally do not apply
to the viscously heated interior regions of the protoplanetary disc where
$\zeta$ increases and the aspect ratio becomes relatively constant
\citep{Bitsch+etal2015a,Ida+etal2016,Brasser+etal2017}.
\begin{figure}
  \begin{center}
    \includegraphics[width=0.9\linewidth]{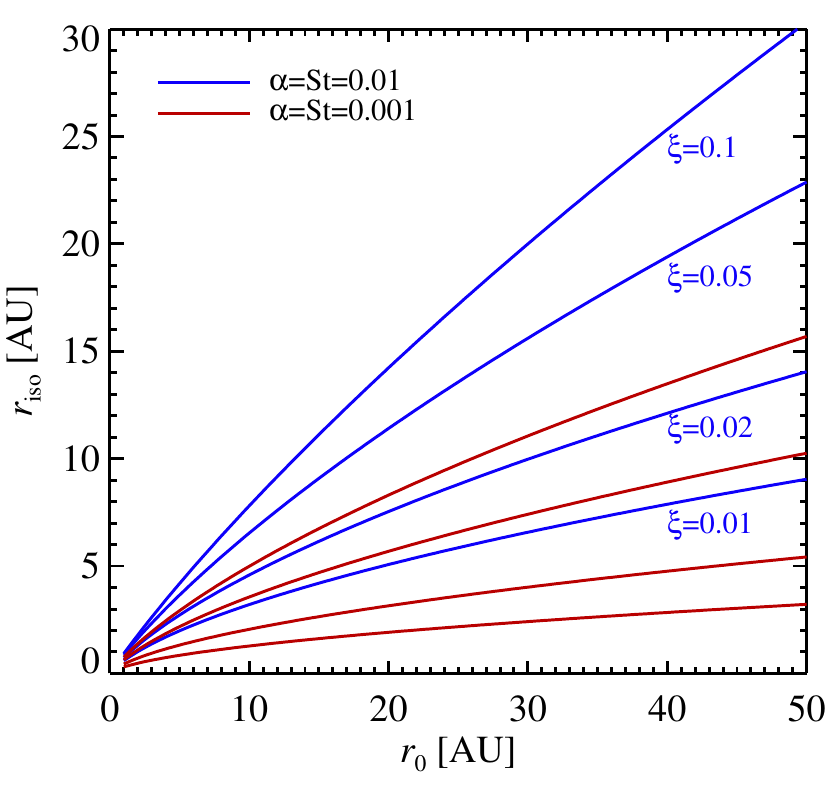}
  \end{center}
  \caption{Radial location of reaching the pebble isolation mass, $r_{\rm iso}$,
  as a function of the starting position in the disc, $r_0$, for different
  values of the pebble-to-gas flux ratio $\xi$. For low values of $\xi$,
  reaching the pebble isolation mass in the 5--10 AU region requires migration
  over tens of AU. Nearly in-situ assembly of a planetary core is possible when
  the pebble-to-gas flux ratio $\xi \gtrsim 0.05-0.1$.}
  \label{f:riso_r0}
\end{figure}

In the limit $A \gg 1$, corresponding to $r_0 \gg r_{\rm iso}$ through equation
(\ref{eq:Miso0}), equation (\ref{eq:X}) has the limiting solution
\begin{equation}
  X = A^{-1/2}
\end{equation}
and hence the isolation mass is reached at radius
\begin{equation}
  r_{\rm iso}/r_0 = A^{-(1/2)[1/(1-\zeta)]} \, .
\end{equation}
We can furthermore make use of $M_{\rm iso,0} \propto r_0^{(3/2)(1-\zeta)}$ and
$M_{\rm max} \propto r_0^{(3/4)(1-\zeta)}$ to infer $A =(M_{\rm iso,0}/M_{\rm
max})^{4/3} \propto r_0^{1-\zeta}$. That gives now the simple relation
\begin{equation}
  r_{\rm iso} \propto r_0^{1/2} \, .
\end{equation}
The length over which the protoplanet migrates before reaching pebble isolation
mass is therefore a steeply increasing function of the starting position. In the
same limit ($r_0 \gg r_{\rm iso}$) the reached pebble isolation mass becomes
simply $M_{\rm iso}=M_{\rm max}$, by inserting $r_{\rm iso}/r_0$ in equation
(\ref{eq:Miso0}). This simple result arises because the growth track turns
nearly horizontal after a significant migration distance and hence the core is
close to its maximum mass $M_{\rm max}$ when it finally reaches pebble isolation
mass.

In Figure \ref{f:riso_r0} we plot the calculated distance of reaching isolation
mass, $r_{\rm iso}$, as a function of the starting position, $r_0$, for
different values of $\alpha={\rm St}$ and $\xi$. The distance is obtained
from the full solution of equation (\ref{eq:pol}). For a core to reach isolation
mass in the 5--10 AU region, the protoplanet must generally start beyond 20 AU,
for nominal values of $\xi=0.01-0.02$. Nearly in-situ core assembly by pebble
accretion requires much higher values of $\xi \gtrsim 0.05-0.1$.

\section{Growth tracks including gas accretion}
\label{s:full_growth}

We now include the effect of gap formation and gas accretion on planetary growth
tracks. We turn to numerical integration, since the reduction of the migration
rate and gas accretion rate by gap formation render the governing equations much
more complex than in the previous section.

\subsection{Type I migration and relative gap height}

\cite{Kanagawa+etal2018} performed a suite of 2-D simulations to measure the
torque on embedded planets of a wide range of masses. They found that the torque
is well-described by the classical Type I torque, which gives rise to the
migration rate expression given in equation (\ref{eq:rdot}), multiplied by the
relative gap height,
\begin{equation}
  \varGamma = -c_{\rm mig} \varGamma_0  \frac{\varSigma_{\rm gap}}{\varSigma_{\rm g}} \, .
\end{equation}
Here $c_{\rm mig}=k_{\rm mig}/2$ is the torque prefactor, $\varSigma_{\rm gap}$
is the surface density in the gap, $\varSigma_{\rm g}$ is the unperturbed
surface density and $\varGamma_0$ ($\propto M^2$) is the natural torque scaling.
The surface density at the bottom of the gap is fitted well by the expression
\citep{DuffellMacFadyen2013,Fung+etal2014,Kanagawa+etal2015,FungChiang2016}
\begin{equation}
  \frac{\varSigma_{\rm gap}}{\varSigma_{\rm g}} = \frac{1}{1+0.04 K} \, ,
  \label{eq:Smin}
\end{equation}
where
\begin{equation}
  K = \left( \frac{M}{M_\star} \right)^2 \left( \frac{H}{r} \right)^{-5}
  \alpha_{\rm v}^{-1} \, .
  \label{eq:K}
\end{equation}
This implies that the migration rate, $\dot{r} = 2 \varGamma/(M v_{\rm K})$,
falls as $1/M$ above the gap transition mass $M_{\rm gap}$ (defined as the mass
for which $K=1/0.04$). The gap transition mass and the pebble isolation mass are
actually closely related concepts. While the gap transition mass measures the
mass required to make a relative gap height of 0.5, the pebble isolation mass
measures the first appearance of a point of zero pressure gradient at the
outside of the planetary gap. The latter criterion is slightly easier to
fulfill, as a relative gap height of around 0.85 is sufficient to invert the
pressure gradient in the simulations of \cite{Bitsch+etal2018b} (we inferred
this relative gap height from their Figure A.1).

\begin{figure}
  \begin{center}
    \includegraphics[width=0.9\linewidth]{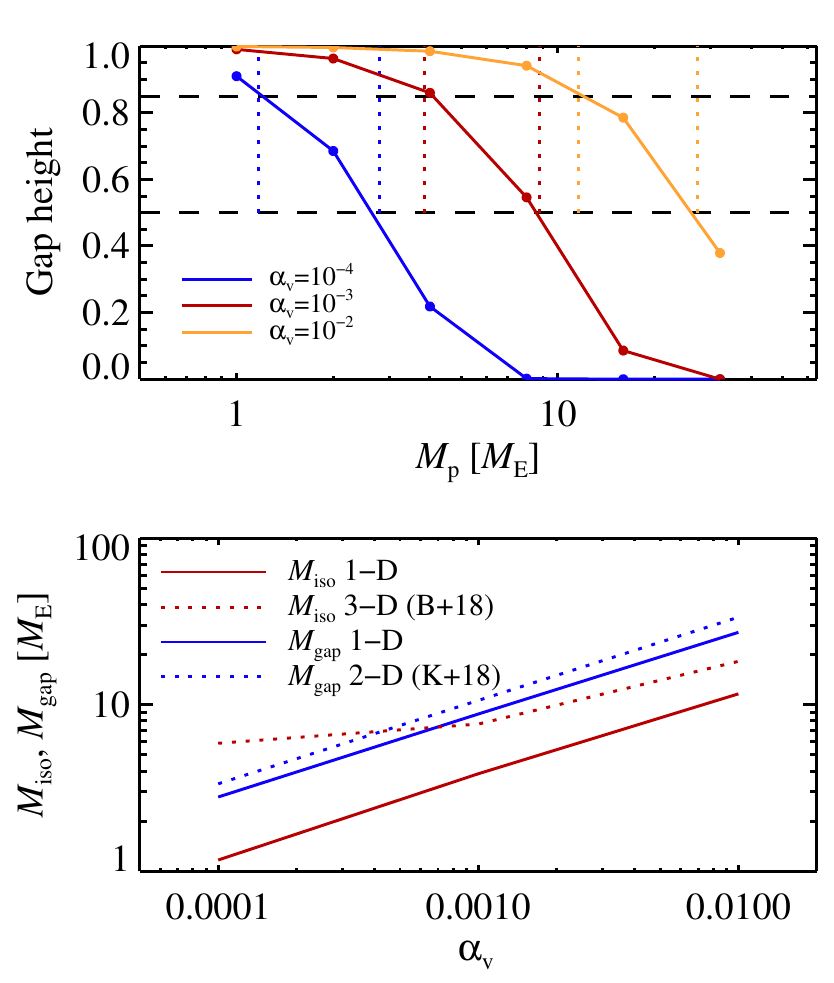}
  \end{center}
  \caption{The results of 1-D simulations of an accretion disc with an embedded
  planet whose torque on the gas is based on the parametrisations of
  \cite{DAngeloLubow2010}. The top panel shows the equilibrium relative gap
  height as a function of the planetary mass, for three values of the viscous
  $\alpha_{\rm v}$. The dashed lines mark the 50\% relative gap height and
  the 85\% relative gap height, the latter approximately corresponding to
  the pebble isolation mass. The dotted lines show the measured pebble
  isolation mass and 50\% gap mass. The bottom panel shows the pebble isolation
  mass ($M_{\rm iso}$) and the gap transition mass ($M_{\rm gap}$) as a function
  of $\alpha_{\rm v}$. We find generally that $M_{\rm gap} \approx 2.3 M_{\rm
  iso}$. The measured gap transition mass corresponds well to the 2-D
  simulations \cite{Kanagawa+etal2018}, while the measured pebble isolation mass
  is about a factor two lower than reported in \cite{Bitsch+etal2018b} and
  displays a more consistent drop with lower $\alpha_{\rm v}$.}
  \label{f:PIM_GTM_Mp}
\end{figure}

To check the robustness of the relative gap height needed for pebble isolation,
we performed additional 1-D simulations of an accretion disc with an embedded
planet, varying the turbulent viscosity. The torque from the planet was mimicked
using the torque profile of \cite{DAngeloLubow2010}, based on their 3-D
simulations. The simulations were run until the relative gap height reached
equilibrium between the gap-opening torque and the viscous momentum transport
($10^3$, $10^4$, $10^5$ yr for $\alpha_{\rm v}=10^{-2}$, $10^{-3}$, $10^{-4}$,
respectively). The results are shown in Figure \ref{f:PIM_GTM_Mp} and confirm
that the pebble isolation mass is reached at a relative gap height of 0.85. We
find a general scaling that the gap transition mass, i.e.\ the mass of 50\%
relative gap height, is 2.3 times the pebble isolation mass. The measured gap
transition mass agrees well with the relative gap height scaling in equations
(\ref{eq:Smin}) and (\ref{eq:K}), particularly that the gap transition mass
scales with the square root of $\alpha_{\rm v}$.

We nevertheless encountered some discrepancy between the $\alpha_{\rm
v}$-dependence of the pebble isolation mass inferred from our simulations and
those of \cite{Bitsch+etal2018b}. At high and medium $\alpha_{\rm v}$ ($10^{-2}$
and $10^{-3}$) our 1-D pebble isolation mass is approximately a factor two lower
than the 3-D pebble isolation mass of \cite{Bitsch+etal2018b}. It is
nevertheless well-known that gaps formed by a 1-D torque prescription become
artificially deeper than what is found in 2-D and 3-D simulations
\citep{HallamPaardekooper2017}. For low $\alpha_{\rm v}$ ($10^{-4}$) the pebble
isolation mass of by \cite{Bitsch+etal2018b} furthermore showed a weak
logarithmic scaling with $\alpha_{\rm v}$ (equation \ref{eq:Miso}), while our
1-D pebble isolation mass maintains its proportionally to the square root of
$\alpha_{\rm v}$, similar to the scaling of the gap transition mass with
$\alpha_{\rm v}$ (equations \ref{eq:Smin} and \ref{eq:K}). This difference could
be due to (1) Rossby wave instabilities triggered at the gap edge at low
$\alpha_{\rm v}$ in 3-D simulations or (2) that the simulations of
\cite{Bitsch+etal2018b} were run only for 1,000 orbital time-scales at the
planet position. The 2-D simulations of \cite{Ataiee+etal2018} also show a
somewhat weaker dependence of the pebble isolation mass on $\alpha_{\rm v}$ than
the square root dependence that we find in Figure \ref{f:PIM_GTM_Mp}.
Identifying the actual reason for the discrepancy will require a dedicated study
beyond the scope of the current paper. In our nominal model we will
therefore set the pebble isolation mass according to \cite{Bitsch+etal2018b}
and the gap transition mass to 2.3 times the pebble isolation mass. This yields
a modified migration equation
\begin{equation}
  \dot{r} = \frac{\dot{r}_{\rm I}}{1+[M/(2.3 M_{\rm iso})]^2} \, ,
\end{equation}
where $\dot{r}_{\rm I}$ is the classical Type I migration rate and $M_{\rm iso}$
is the pebble isolation mass given in equation (\ref{eq:Miso}). We will
also show results in Section \ref{s:variation} where we (a) use $M_{\rm gap}$
from the 2-D simulations of \cite{Kanagawa+etal2018} and $M_{\rm iso}$ from the
3-D simulations of \cite{Bitsch+etal2018b} or (b) use $M_{\rm gap}$ from the 2-D
simulations and $M_{\rm iso}=M_{\rm gap}/2.3$.

Reaching first the pebble isolation mass and shortly after the gap transition
mass thus signify three important mile-stones for the growth of a protoplanet:
(a) the end of the accretion of pebbles, (b) the beginning of gas contraction
and (c) the transition to a slow-down in the migration caused by the deepening
gap.
\begin{figure*}
  \begin{center}
    \includegraphics[width=0.9\linewidth]{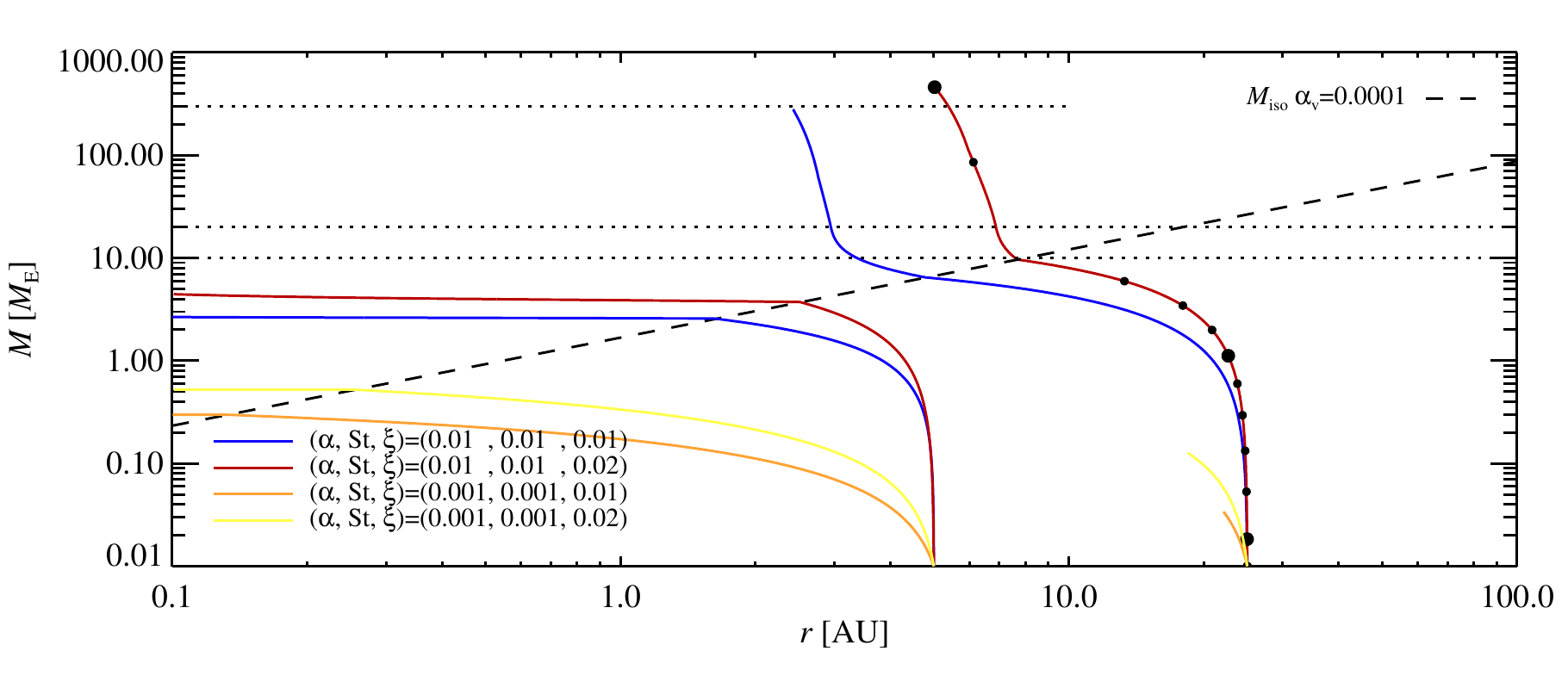}
  \end{center}
  \caption{Numerically integrated growth tracks for core accretion to the pebble
  isolation mass followed by gas accretion. We start the protoplanets at a mass
  of $M_0=0.01\,M_{\rm E}$ at $t_0=0.9\,{\rm Myr}$ (except for the $\alpha={\rm
  St}=0.01$ growth track starting from 25 AU, which we start at $t_0=0.3\,{\rm
  Myr}$) in a protoplanetary disc that evolves over a total time of 3 Myr. The
  dots indicate the time for the growth track that successfully forms a Jupiter
  analogue (small dots are separated by 0.2 Myr, large dots indicate a time of
  3, 2 and 1 Myr). We use a migration model where the migration rate is reduced
  by multiplication with the relative gap height, following
  \cite{Kanagawa+etal2018}. The migration rate is therefore inversely
  proportional to the planet mass beyond the pebble isolation mass. This results
  in gas accretion over a few astronomical units of migration.}
  \label{f:growth_tracks_numerical}
\end{figure*}

\subsection{Gas accretion}

The gas accretion rate of a protoplanet embedded in a protoplanetary disc is
highly uncertain. Even the term {\it accretion} is, in our opinion, slightly
misleading, as the mass increase takes place in the form of a contraction of the
envelope due to entropy reduction by radiative heat loss. \cite{KlahrKley2006}
demonstrated that the circumplanetary disc formed in some isothermal simulations
is replaced by a hot, hydrostatic gas blob in simulations including compressive
heating and radiative transfer. \cite{Ikoma+etal2000} performed 1-D simulations
of the envelope contraction and found run-away contraction where the energy
loss, and hence contraction rate, is accelerated at higher envelope masses.
\cite{DAngeloBodenheimer2013} presented extensive simulations of the gas
contraction onto low-mass cores and found, as in earlier 1-D work, that the
contraction is accelerated at higher core mass \citep[the same trend was
observed in][]{LambrechtsLega2017}. \cite{Lambrechts+etal2018} measured the gas
contraction rate in hydrodynamical simulations with radiative transfer, for
planetary cores of up to Jupiter-mass. They found that the accretion rate is
orders of magnitude lower than the mass flux passing through the Hill radius, as
most of this gas is transported out again on horse-shoe orbits and more complex
streamlines.

We use here the gas accretion prescription proposed in \cite{Ida+etal2018},
taking into account both the Kelvin-Helmholtz-like contraction of the envelope
and the feeding of gas from the protoplanetary disc into the Hill sphere of the
protoplanet. The contraction of the gaseous envelope is assumed to commence
after the core reaches pebble isolation mass, at a rate motivated by
\cite{Ikoma+etal2000},
\begin{equation}
  \left( \frac{\de M}{\de t} \right)_{\rm KH} = 10^{-5}\,M_{\rm E}\,{\rm
  yr}^{-1} \left( \frac{M}{10 M_{\rm E}} \right)^4 \left(
  \frac{\kappa}{0.1\,{\rm m^2\,kg^{-1}}} \right)^{-1} \, .
  \label{eq:dMdtKH}
\end{equation}
Here $\kappa$ is the opacity of the envelope, which we discuss further below.
This accretion rate will become larger than what the protoplanetary disc can
supply for high planetary masses. \cite{TanigawaTanaka2016} used isothermal,
global simulations to demonstrate that gas enters the Hill sphere at a rate
\begin{eqnarray}
  \left( \frac{\de M}{\de t} \right)_{\rm disc} &=& \frac{0.29}{3 \pi} \left(
  \frac{H}{r} \right)^{-4} \left( \frac{M}{M_\star} \right)^{4/3}
  \frac{\dot{M}_{\rm g}}{\alpha} \frac{\varSigma_{\rm gap}}{\varSigma_{\rm g}} \nonumber \\ &=& 1.5 \times 10^{-3}\,M_{\rm
  E}\,{\rm yr}^{-1} \left( \frac{H/r}{0.05} \right)^{-4} \left(
  \frac{M}{10\,M_{\rm E}} \right)^{4/3} \nonumber \\
  & & \times \left( \frac{\alpha}{0.01} \right)^{-1} \left( \frac{\dot{M}_{\rm
  g}}{10^{-8}\,M_\odot\,{\rm yr}^{-1}} \right) \frac{1}{1+(M/M_{\rm gap})^2} \, .
  \label{eq:dMdtdisc}
\end{eqnarray}
This equation is derived in the companion paper by \cite{Ida+etal2018} -- we
note here that the simpler form, $(\de M/\de t)_{\rm disc} \propto R_{\rm H}^2
\varOmega \varSigma_{\rm gap} (R_{\rm H}/H)^2 $, shows that the equation can be
decomposed into the full flux into the Hill sphere multiplied by the squared
ratio of the Hill radius to the scale-height. The latter reflects an
empirical finding that that the protoplanet increases its ability to accrete the
gas streamlines that enter the Hill radius as the mass increases
\citep{TanigawaTanaka2016}.

The prefactor in equation (\ref{eq:dMdtdisc}) can be converted to solar units to
compare with the disc accretion rate, $1.5 \times 10^{-3}\,M_{\rm E}\,{\rm
yr}^{-1} \approx 4.5 \times 10^{-9}\,M_\odot\,{\rm yr}^{-1}$. If the planet
contracts rapidly enough to absorb the entire gas flow into the Hill radius,
then we must therefore limit the accretion rate to the global gas accretion rate
\citep[see][for a discussion of this limit]{LubowDAngelo2006}. We therefore set
the gas accretion rate of the planet equal to
\begin{equation}
  \left( \frac{\de M}{\de t} \right)_{\rm g} = {\rm min}\left[ \left( \frac{\de
  M}{\de t} \right)_{\rm KH}, \left( \frac{\de M}{\de t} \right)_{\rm disc},
  \dot{M}_{\rm g} \right] \, .
\end{equation}
\cite{Ikoma+etal2000} found that the Kelvin-Helmholtz gas accretion rate
increases inversely proportionally to the opacity $\kappa$ (equation
\ref{eq:dMdtKH}). \cite{BitschJohansen2016} calculated an opacity of
micron-sized ice particles in the range $0.01-0.1\,{\rm m^2\,kg^{-1}}$.  For
core accretion by pebble accretion, we expect that the accreted gas will be
completely pebble-free beyond the pebble isolation mass
\citep{Lambrechts+etal2014}. Small dust can nevertheless pass the planet barrier
together with the accreted gas, constituting maybe 10\% of the total content of
solids. That gives a nominal opacity in the range $0.001-0.01\,{\rm
m^2\,kg^{-1}}$. The opacity could be even lower if the grains in the envelope
coagulate and sediment \citep{Mordasini2014}. We adopt here a standard opacity
value of $\kappa = 0.005 \,{\rm m^2\,kg^{-1}}$.
\begin{figure*}
  \begin{center}
    \includegraphics[width=\linewidth]{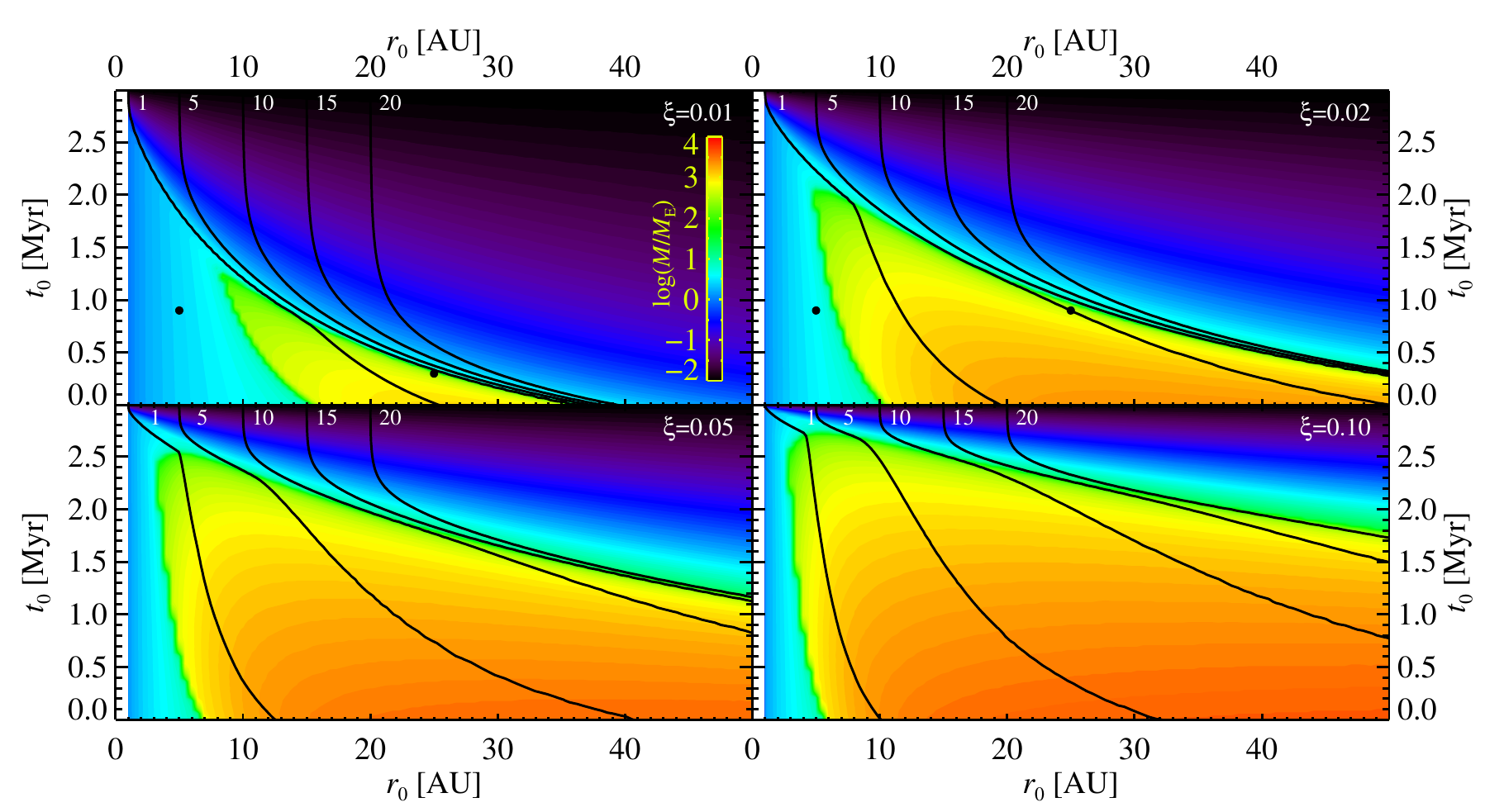}
  \end{center}
  \caption{Growth maps showing final masses (colors) and selected final
  positions (black contours) of protoplanets starting at 1 to 50 AU distance
  from the star after between 0 and 3 Myr of disc evolution. The four panels
  show the results for different values of the pebble-to-gas flux ratio $\xi$.
  The starting points of the selected growth tracks from Figure
  \ref{f:growth_tracks_numerical} are indicated with dots. The cores of cold gas
  giants akin to Jupiter and Saturn start their assembly here in the 20--30 AU
  region for nominal values of $\xi=0.01-0.02$.  Higher values of $\xi$ allow
  core assembly closer to the central star, starting in the 10--15 AU region.}
  \label{f:growth_maps}
\end{figure*}

\subsection{Protoplanetary disc model}

The growth rates reported in \cite{Ikoma+etal2000} were measured at a single
value of the gas surface density and gas temperature at the outer boundary.
Hence the dependence of the accretion rates on the gas surface density and
temperature is not known. If the gas contraction is indeed limited by the
ability of the envelope to cool, then the outer boundary condition may not
matter much \citep[as found by][]{PisoYoudin2014,LeeChiang2015}. This, in
turn, implies that mass accretion wins more easily over migration when the gas
disc is depleted compared to the primordial value, since the initial gas
accretion rate is relatively unaffected by the protoplanetary disc surface
density, while the migration rate is proportional to the surface
density\footnote{The reduced migration rate at low gas column densities is
an important component of the model of super-Earth formation in depleted gas
discs presented in \cite{LeeChiang2016}.}.

The dependence of the migration rate on the gas surface density, combined with
the non-dependence of the Kelvin-Helmholtz contraction rate on the surface
density, therefore necessitates a specific disc evolution model to integrate the
growth tracks. We adopt here a nominal viscous accretion disc model where
$\dot{M}_{\rm g}$ evolves from $10^{-7}\,M_\odot\,{\rm yr}^{-1}$ to
$10^{-8}\,M_\odot\,{\rm yr}^{-1}$ over 3 Myr \citep[typical for solar-mass
stars, see][]{Hartmann+etal2016}. Following
\cite{Hartmann+etal1998}, the accretion rate onto the star evolves as
\begin{equation}
  \dot{M}_{\rm g}(t) = \dot{M}_0 \left[ \frac{t}{t_{\rm s}} + 1
  \right]^{-(5/2-\gamma)/(2-\gamma)} \, ,
\end{equation}
where $\gamma$ is the power-law index of the turbulent viscosity, $\nu \propto
r^\gamma$, and the characteristic time-scale, $t_{\rm s}$, is
\begin{equation}
  t_{\rm s} = \frac{1}{3(2-\gamma)^2} \frac{R_1^2}{\nu_1} \, .
\end{equation}
Here $R_1$ is the initial characteristic disc size and $\nu_1$ is the value of
the viscosity at that radius. Knowing the temperature power law index $\zeta$
results in $\gamma=3/2-\zeta$ ($=15/14$ for $\zeta=3/7$). We proceed by choosing
$R_1$ to yield the desired starting and ending mass accretion rate over an
assumed disc life-time of 3 Myr. In Section \ref{s:ppd_sizes} we present
planet formation results for protoplanetary discs with initially smaller sizes
and hence a steeper temporal decay of their mass accretion rates.

We take into account in the numerical simulations that the pebble accretion rate
is lowered when the pebble scale-height is higher than the pebble accretion
radius, following the method described in \cite{Johansen+etal2015}.
\cite{JohansenKlahr2005} demonstrated in MHD simulations that the turbulent
viscosity $\alpha_{\rm v}$ and the turbulent diffusion coefficient $\delta$ are
approximately equal; we therefore for simplicity assume $\delta=\alpha_{\rm v}$
in our simulations.
\begin{figure*}
  \begin{center}
    \includegraphics[width=\linewidth]{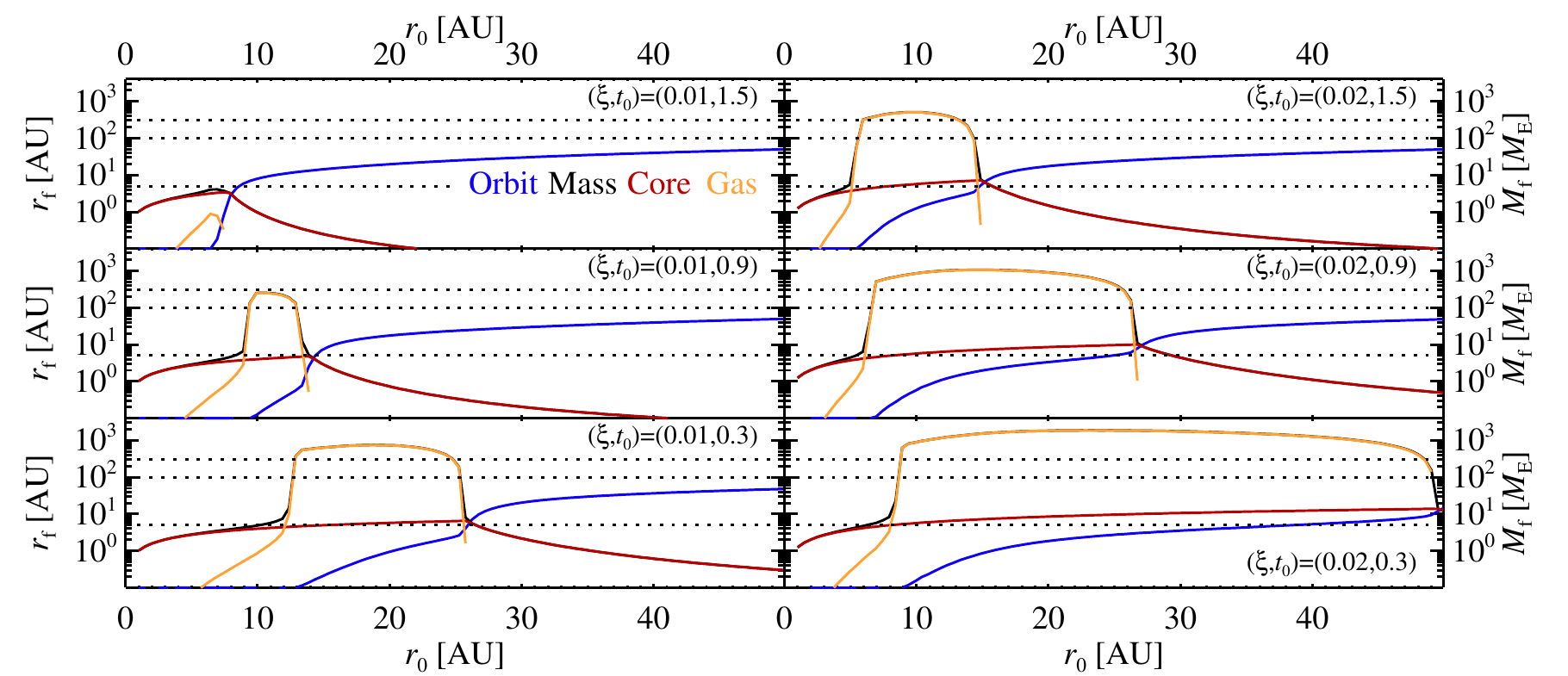}
  \end{center}
  \caption{The final orbits (left axis) and final total mass, core mass and gas
  (right axis) for protoplanets starting from 1 to 50 AU (bottom/top axis), at
  three starting different times, $t_0$=0.3,0.9,1.5 Myr (bottom to top panels)
  and for two values of $\xi$=0.01,0.02 (left and right panels). The dotted
  lines indicate masses or orbits of 5, 100 and 300. Protoplanets starting far
  from the star experience both little growth and migration. As the starting
  position approaches the star, the core mass grows towards 10\,$M_{\rm E}$,
  triggering rapid migration and gas accretion. However, protoplanets that start
  interior of 5 AU have very protracted gas accretion due to the low pebble
  isolation mass there. The best analogues of Jupiter, Saturn, Uranus and
  Neptune start their assembly in the region around 25 AU for $\xi=0.02$ and
  $t_0=0.9$ Myr.}
  \label{f:rf_Mf_r0}
\end{figure*}

\subsection{Growth tracks with gas accretion}

The integration of the growth track bundles from Figure
\ref{f:growth_tracks_analytical} including gas accretion are shown in Figure
\ref{f:growth_tracks_numerical}. We start the growth tracks at $t_0=0.9$ Myr and
integrate until the disc is assumed to dissipate after $t=3$ Myr. To demonstrate
the effect of the starting time, we start one growth track (with $\alpha={\rm
St}=0.01$ and $\xi=0.01$ starting at $r_0=25\,{\rm AU}$) at $t_0=0.3\,{\rm
Myr}$. The core growth tracks before reaching pebble isolation mass are
relatively unaffected by the slow-down of the migration rate towards the gap
transition mass (and follows the analytical solution that we derived in Section
\ref{s:core_growth}, even when we include here that the initial pebble accretion
is reduced in 3-D). However, after reaching the pebble isolation mass, the two
growth tracks starting at 25 AU with $\alpha={\rm St}=0.01$ turn quickly
upwards, as the migration rate falls inversely proportional to the planetary
mass. This results in gas accretion length scales that are much shorter than the
core accretion length scale. In contrast to this, the growth tracks starting at
5 AU reach an isolation mass of $M_{\rm iso} \sim 1\ldots5\,{\rm M_{\rm E}}$. At
those core masses, Kelvin-Helmholtz contraction is very slow, according to
equation (\ref{eq:dMdtKH}), and these planets migrate towards the star with
little success at accreting gas.

In Figure \ref{f:growth_maps} we show the final positions of protoplanets
starting at 1 to 50 AU distance from the star in the protoplanetary disc and
starting times between 0 and 3 Myr, for four values of the pebble-to-gas flux
ratio $\xi$. The slow-down of migration in the gas accretion phase allows more
space for migration during the core accretion stage, hence we can form gas
giants at a lower pebble accretion rate than in \cite{Bitsch+etal2015b}. For
nominal pebble-to-gas flux ratios of $\xi=0.01-0.02$, the cores of cold gas
giants with final orbits in the 5--10 AU region (analogues of Jupiter and Saturn
in the Solar System) start their assembly much further out, in the 20--30 AU
region. Shorter migration distances, with starting locations in the 10--15 AU
region, are possible at higher pebble-to-gas flux ratios $\xi=0.05-0.1$.

We find generally that our integrations produce much fewer hot and warm gas
giants, with final orbits interior to 1 AU, compared to simulations that
considered the traditional Type II migration
\citep{BitschJohansen2016,BitschJohansen2017}. This is due to the slow-down of
the migration rate with the formation of a deep gap when using the
\cite{Kanagawa+etal2018} migration prescription here. Other studies have invoked
photoevaporation as an effect to prevent massive planets from migrating all the
way to the inner edge of the protoplanetary disc
\citep{AlexanderPascucci2012,ErcolanoRosotti2015}; here and in
\cite{Ida+etal2018} we show that massive planets may be naturally prevented from
migrating to the disc edge due to the deep gas gaps that they carve.

The core masses of our synthetic planets are nevertheless smaller than what is
inferred for total content of heavy elements in the gas giants in the Solar
System. The recent data on the gravitational field of Jupiter by the Juno
satellite yield a total amount of heavy elements between $25\,M_{\rm E}$ and
$45\,M_{\rm E}$ \citep{Wahl+etal2017}. However, we did not include in our
simulations the contribution from planetesimal pollution during the gas
accretion phase. \cite{ShiraishiIda2008} modelled the pollution by late
planetesimal infall numerically and found accretion of up to $10\,M_{\rm E}$ of
additional planetesimals, bringing the total amount of heavy elements more in
line with observations.

The final orbits and masses for three selected starting times are shown in
Figure \ref{f:rf_Mf_r0}. One sees how the outcome of planet formation can be
broadly divided into three categories. Protoplanets starting their growth far
from the star experience both little growth and little migration and end as ice
planets with masses broadly between 0.1 and 2 $M_{\rm E}$ \citep[as also found
by][]{Bitsch+etal2015a}. As the starting position is reduced, the cores grow
towards the pebble isolation mass of $\approx 10\,M_{\rm E}$. This triggers
rapid gas accretion and the formation of Jupiter and Saturn analogues. The total
gas mass plateaus at around $1000\,M_{\rm E}$; this is an effect of gap
formation that limits gas accretion in our gas accretion model that follows
\cite{TanigawaTanaka2016}. Protoplanets that start closer than approximately 5
AU of the host star have low pebble isolation masses and hence never experience
gas accretion. These form analogues of the super-Earths found around many stars
\citep{Batalha+etal2013}, although we emphasize here that the mutual dynamics of
trains of such migrating super-Earths, not included here, is key to
understanding the final orbits and multiplicities of super-Earth-mass migrators
\citep{Izidoro+etal2017,Matsumura+etal2017}. Alternatively, the orbital
properties of observed super-Earths has been proposed to be consistent with a
lack of planetary migration \citep{LeeChiang2017}. We also ignore here the
truncation of the inner disc by the magnetospheric cavity, which may play an
important role in reducing or halting migration \citep{Liu+etal2017}.  Our model
does not readily explain how to stop the accretion of gas at the mass of Saturn
or Jupiter, unless we invoke special timing of the dissipation of the gaseous
protoplanetary disc. A possible solution to this problem is that the turbulent
viscosity, $\alpha_{\rm v}$, is even lower than what we have assumed and that
both the pebble isolation mass and gap transition mass are lower in real
protoplanetary discs. An earlier onset of gap formation would be an efficient
way to limit the flow of gas to the protoplanet, through equation
(\ref{eq:dMdtdisc}).
\begin{figure}
  \begin{center}
    \includegraphics[type=pdf,ext=.pdf,read=.pdf,width=0.9\linewidth]{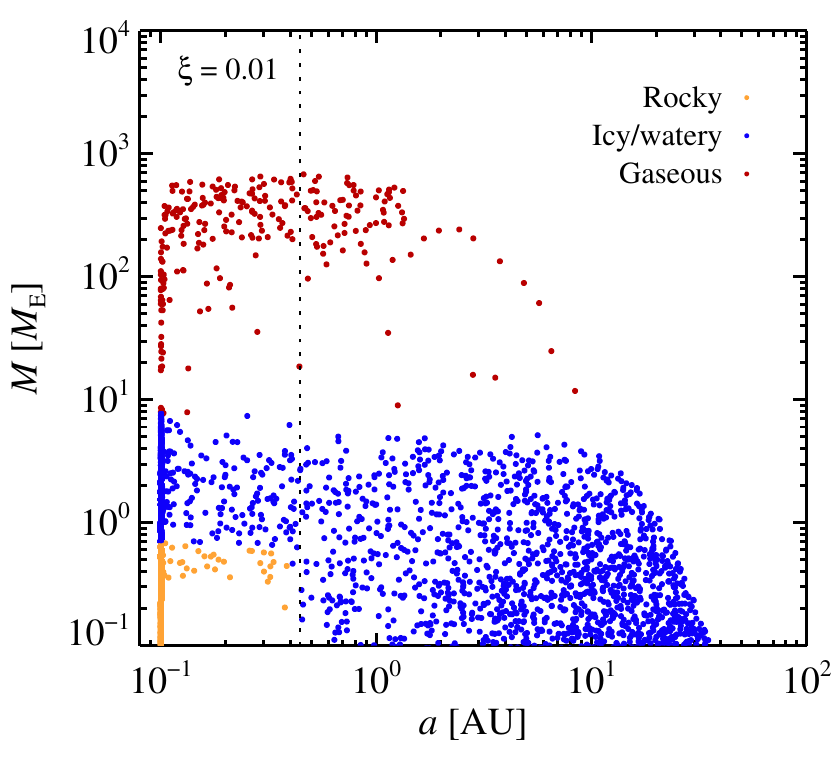}
    \includegraphics[type=pdf,ext=.pdf,read=.pdf,width=0.9\linewidth]{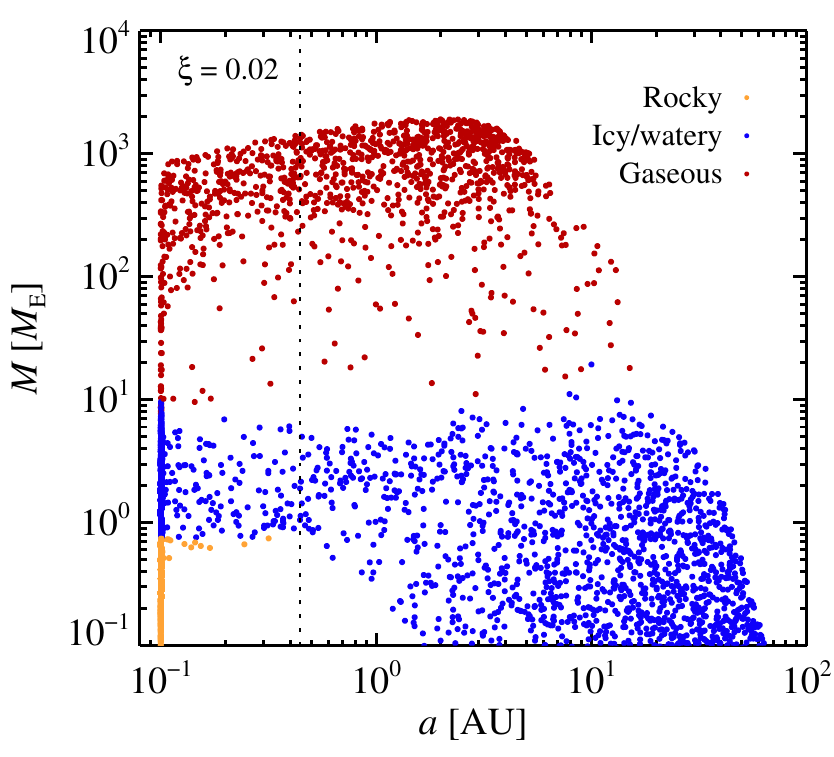}
  \end{center}
  \caption{Population synthesis of two of the growth maps shown in Figure
  \ref{f:growth_maps}. We have selected 10,000 points sampled randomly in the
  logarithmic starting position and the linear starting time. Core-dominated
  planets with an ice or water fraction of less than 25\% are marked in yellow,
  core-dominated planets with an ice or water fraction between 25\% and 50\% are
  marked in blue and gas-dominated planets are marked in red. The water ice line
  is indicated by a dotted line.}
  \label{f:population_synthesis}
\end{figure}

\subsection{Formation of ice giants}

Forming planets with core masses equivalent to those of Uranus and Neptune in
10-20 AU orbits \citep[as is broadly consistent with the initial condition for
the later planetesimal-driven migration of the giant planets][]{Gomes+etal2005}
is possible when a protoplanet starts slightly exterior to those that end up
forming gas giants. However, the parameter space for forming ice giants is quite
small, since gas accretion is rapid for those core masses. If, on the other
hand, the ice giants formed by giant impacts of a higher number of 5-Earth-mass
cores, as explored in \cite{Izidoro+etal2015}, then the seeds of the ice giants
would be allowed to form even further from the Sun and not be in such a high
risk of accreting substantial amounts of gas. Also, in our work we do not
consider the release of water vapour in the gas envelope. The contraction of an
envelope highly enriched in water vapour could explain the relatively small
amount of hydrogen in the ice giants of the solar system
\citep{Lambrechts+etal2014,VenturiniHelled2017}.
\begin{figure}
  \begin{center}
    \includegraphics[type=pdf,ext=.pdf,read=.pdf,width=0.9\linewidth]{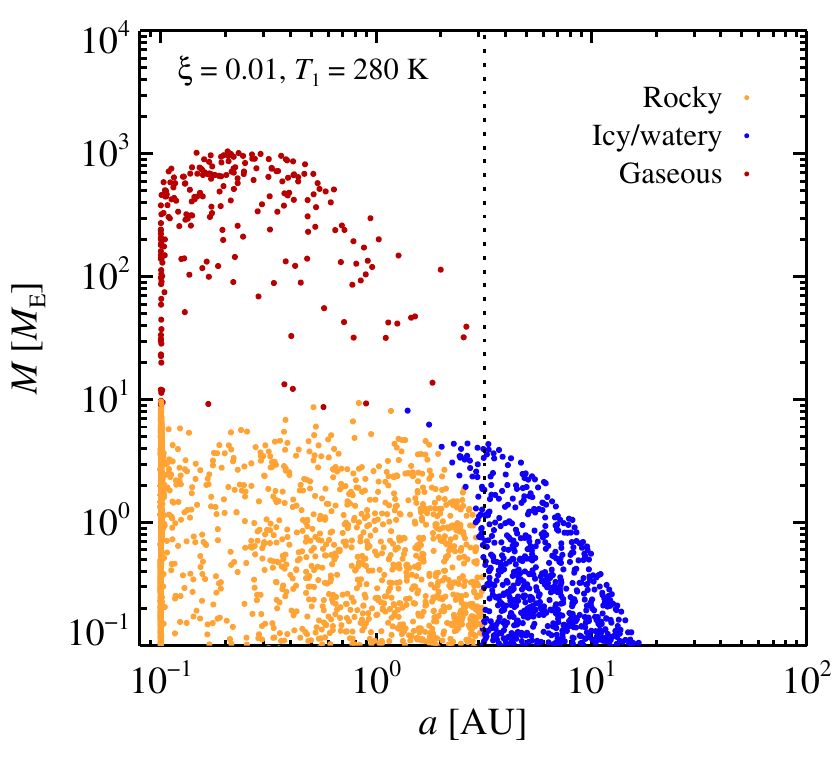}
  \end{center}
  \caption{Population synthesis plot for $\xi=0.01$ and a temperature of
  $T_1=280$ K at $r=1$ AU. The larger aspect ratio $H/r$, compared to Figure
  \ref{f:population_synthesis}, slows down type I migration to a degree that icy
  protoplanets formed beyond the water ice line either stay small or grow into
  gas giant cores. This yields mainly dry super-Earths interior of the water ice
  line, in agreement with constraints from exoplanet observations
  \citep{OwenWu2017}. However, the increased gas scale-height also slows down
  both pebble accretion and gas accretion, resulting in fewer gas giants in the
  outer regions of the protoplanetary disc.}
  \label{f:population_synthesis_T280}
\end{figure}

\subsection{Population synthesis}

In Figure \ref{f:population_synthesis} we illustrate the planetary populations
arising from our models by performing a population synthesis operation on the
growth maps for $\xi=0.01$ and $\xi=0.02$. We have sampled 10,000 random points
in each growth map, using a logarithmic distribution of starting points and a
linear distribution of starting times. We have also calculated the compositions
of the planetary cores, by assuming that half of the mass of the pebbles is in
the form of water ice outside of the water ice line.  The water ice line is
situated at around 0.45 AU in our simulations, due to the fact that we do not
include viscous heating. The population synthesis demonstrates a clear dichotomy
between those planets that reach the pebble isolation mass and undergo rapid gas
accretion and those that stay below the pebble isolation mass. The run-away
nature of early gas accretion leaves the region between 10 and 100 Earth masses
relatively empty \citep[this is known as the planet desert, see][]{IdaLin2004}.
\begin{figure*}
  \begin{center}
    \includegraphics[width=0.9\linewidth]{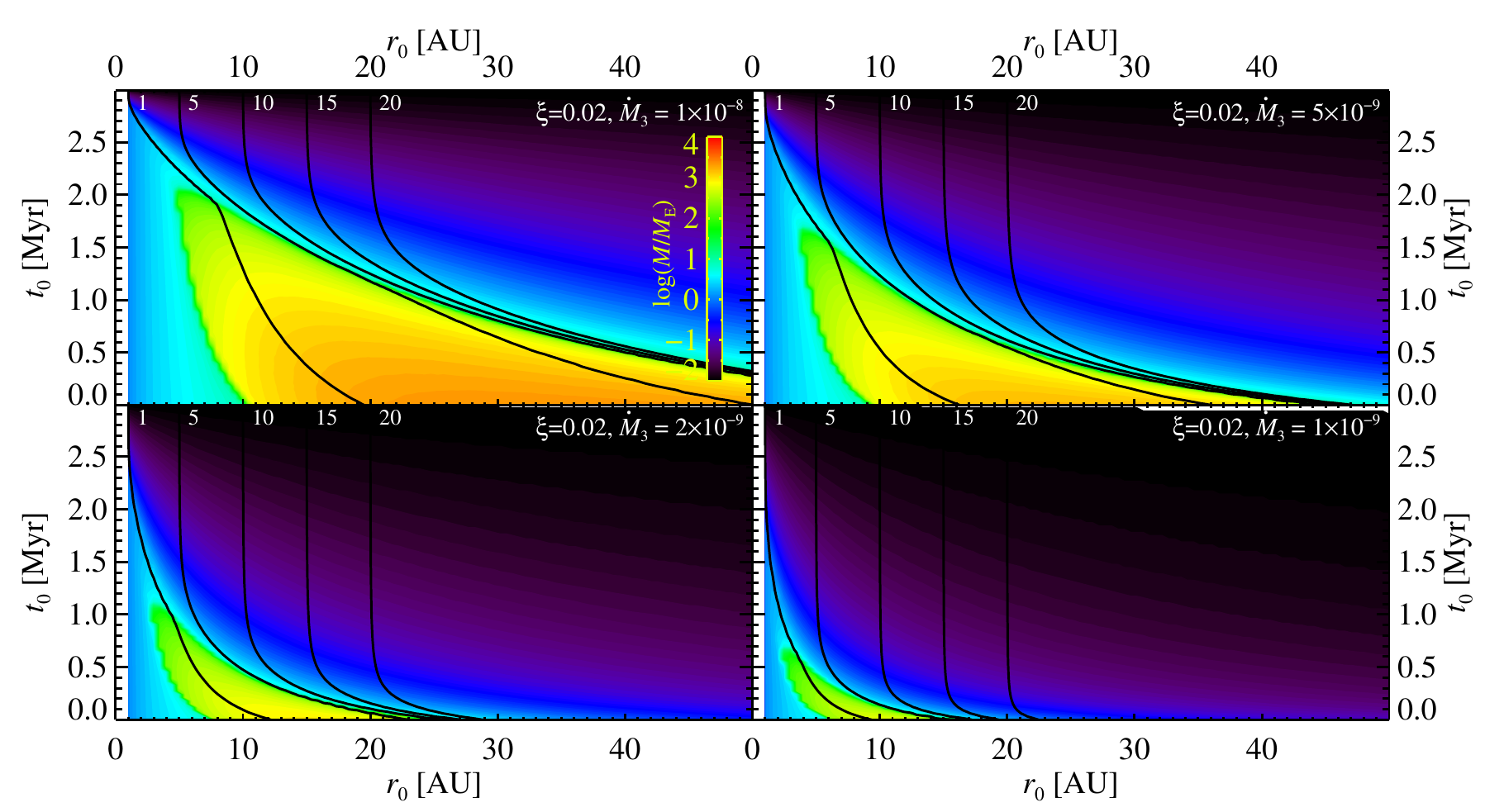}
  \end{center}
  \caption{Growth maps for four values of $\dot{M}_3$, the mass accretion
  rate after 3 Myr, with fixed pebble-to-gas flux ratio $\xi=0.02$. The initial
  accretion rate is $10^{-7}\,M_\odot\,{\rm yr}^{-1}$ in all models. Increasing
  the steepness of the temporal accretion rate profile, obtained here by
  lowering the initial disc size, decreases the region for forming gas-giant
  planets.}
  \label{f:growth_map_Mdot}
\end{figure*}

Our pebble accretion population synthesis plots appear qualitatively similar to
those presented in \cite{BitschJohansen2017} and \cite{Ndugu+etal2018}. We refer
to those two papers for comparisons to the observed exoplanet populations. One
notable difference is the relatively fewer gas giants that reach the inner disc
edge in our simulations. This is mainly an effect of the new planetary migration
prescription that we explore here. Comparing our results to
\cite{Brugger+etal2018} \citep[who used a correction to the pebble sizes and
fluxes of][resulting in very low pebble fluxes]{Bitsch+etal2015b}, we find very
similar synthetic exoplanet populations, although the small-pebble model that we
advocate here requires much lower pebble fluxes (and hence metallicities) to
trigger the formation of gas giants in cold orbits. \cite{Chambers2018}
presented a comprehensive suite of population synthesis calculations and used
the observed exoplanet populations to pin down the best set of model input
parameters for the protoplanetary disc and the pebble sizes. He found best fits
to the data for low diffusion coefficients ($\sim$$10^{-5}$), which can be
understood because of the positive effect of lowering the diffusion coefficient
in the initial 3-D stage of pebble accretion (we discuss this point further in
Section \ref{s:variation}).

\cite{Lin+etal2018} considered planet formation in an $\alpha$-disc model
with an evolving gas and pebble component. They studied a wide range of Stokes
numbers ${\rm St}$ and disc viscosities $\alpha$ and found that it is mainly
possible to form gas-giant planets when ${\rm St}>\alpha$. Smaller Stokes
numbers maintain the global metallicity, as we also show in equation
(\ref{eq:Zp}), but protoplanets accreting small pebbles experience a reduced
pebble accretion rate both in the 2-D and 3-D Hill accretion branch.
\cite{Lin+etal2018} also reported that their synthetic planets either remain
sub-Earth mass or ``explode'' to form gas giants. We recover this result in our
own growth maps when considering only protoplanets started at $t=0$. Looking at
the top left panel of Figure \ref{f:growth_maps}, we see that seeds starting at
$t=0$ in up to 15 AU distances migrate to the inner edge of the disc where they
are stuck at a few Earth masses, while the seeds starting between 15 and 35 AU
grow to warm and cold gas giants. Seeds starting even further out experience
only some minor growth in-situ, due to the long growth time-scales there. In our
work we keep the actual time that it takes to form cores of $M=0.01\,M_{\rm E}$
as a free parameter in our growth maps; the actual time that it takes to grow
from planetesimals to such protoplanets is largely unknown and may be determined
by the interplay between pebble accretion in the inefficient Bondi branch and
mutual planetesimal-planetesimal collisions \citep{JohansenLambrechts2017}.
\begin{figure*}
  \begin{center}
    \includegraphics[width=0.9\linewidth]{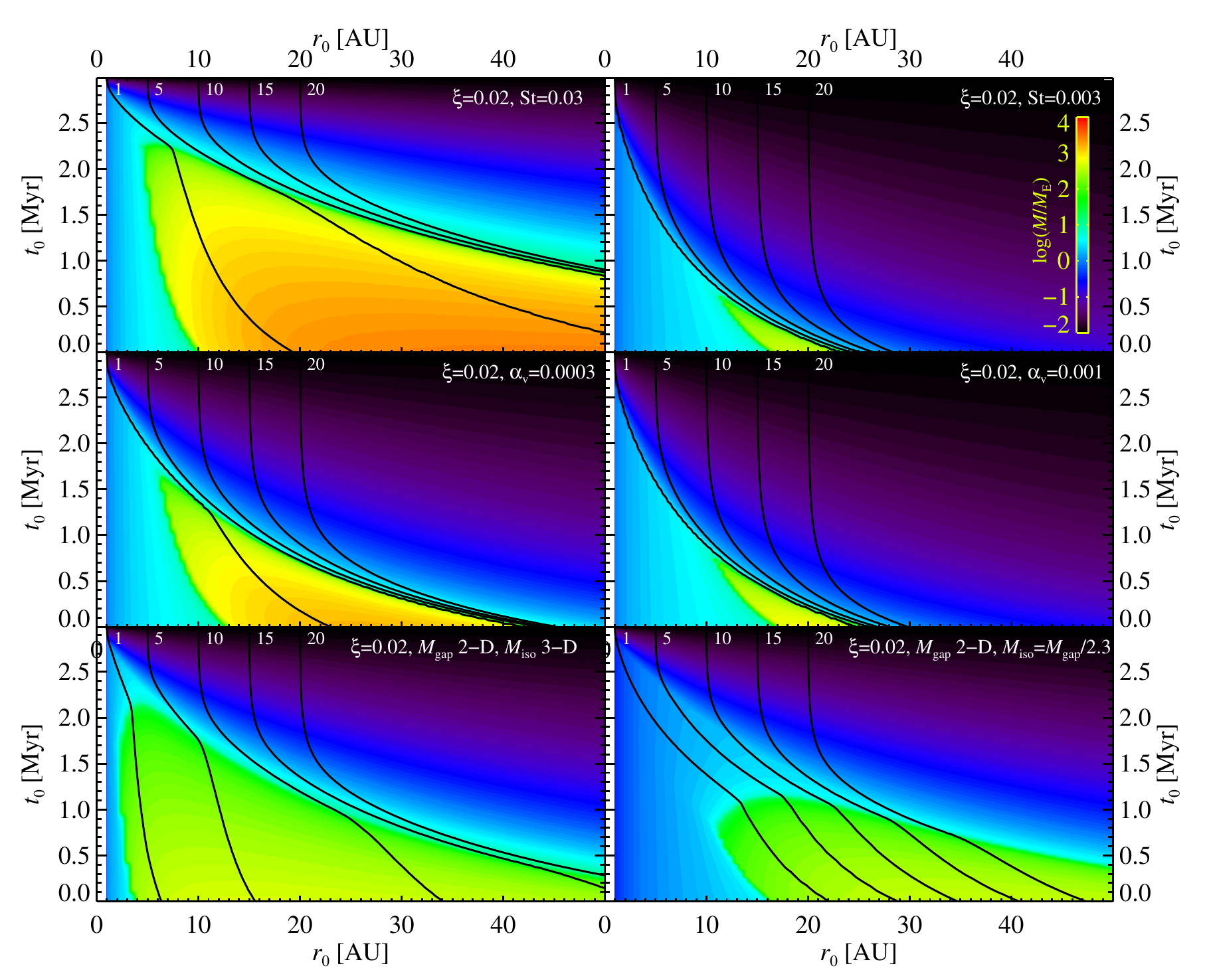}
  \end{center}
  \caption{Growth maps for six variations of our nominal model. {\it Top
  panels}: higher and lower Stokes number, ${\rm St}=0.03$ and ${\rm St}=0.003$.
  The Stokes number is an important factor in the pebble accretion rate,
  particularly in the initial stage of 3-D pebble accretion. {\it Middle
  panels:} Higher turbulent viscosity (and turbulent diffusion), $\alpha_{\rm
  v}=0.0003$ and $\alpha_{\rm v}=0.001$. Planet formation is pushed to earlier
  times as the turbulent viscosity is increased. {\it Bottom panels:} variations
  of the gap formation prescriptions, either using the 50\% gap mass from the
  2-D simulations of \cite{Kanagawa+etal2018} and the pebble isolation mass from
  the 3-D simulations of \cite{Bitsch+etal2018b} ({\it bottom left}) or using
  the 50\% gap mass from the 2-D simulations of \cite{Kanagawa+etal2018} and a
  pebble isolation mass that is a factor 2.3 times below that value. Both result
  in decreased migration and decreased gas accretion.}
  \label{f:growth_map_variation}
\end{figure*}

Super-Earths with final orbits within the water ice line in Figure
\ref{f:population_synthesis} transition from dry to water-rich at a mass of
approximately one Earth mass, as more massive migrators accrete a large fraction
of their mass outside of the water ice line. This composition appears to be in
contrast with the inferred rocky compositions of super-Earths
\citep{OwenWu2017}. However, we have neglected in this paper the viscous heating
that would increase the disc aspect ratio $H/r$ in the inner regions of
protoplanetary discs. An increase in the disc aspect ratio leads to lower
migration rates in the inner protoplanetary disc and hence slows down the icy
protoplanets that penetrate from beyond the ice line. The outwards migration
resulting from such viscous heating will furthermore block the inwards flow of
icy protoplanets of a certain mass range \citep{BitschJohansen2016} -- we
plan to include this effect in future studies using our small Stokes number
model.

In Figure \ref{f:population_synthesis_T280} we show a population synthesis
calculation on a model where the temperature at 1 AU is set to the higher value
$T_1=280$ K, corresponding to a sound speed of $c_{\rm s1}=9.9\times10^2\,{\rm
m\,s^{-1}}$. This temperature profile corresponds to the Minimum Mass Solar
Nebula model of \cite{Hayashi1981}. Through equation (\ref{eq:Mmaxs}) this
increase in sound speed leads to a larger maximum mass, due to the slower
migration speed at an increased $H/r$. The result is that terrestrial planets
and super-Earths interior of the water ice line are now mainly rocky. Icy
protoplanets formed exterior of the ice line either stay small, if they form
late, or grow to gas giants.

\subsection{Different temporal accretion profiles}
\label{s:ppd_sizes}

We have so far analysed planet formation in a protoplanetary disc that evolves
its accretion rate from $10^{-7}\,M_\odot\,{\rm yr}^{-1}$ to
$10^{-8}\,M_\odot\,{\rm yr}^{-1}$ over 3 Myr. Using $\alpha=0.01$ and our
adapted temperature profile, such a temporal evolution implies that the initial
disc size was $R_1=90.8$ AU and contained a mass of $0.12\,M_\odot$ of gas and
400 $M_{\rm E}$ of solids at a metallicity $Z=0.01$. These values are similar to
what is inferred for the iconic protoplanetary disc around HL Tau
\citep{Carrasco-Gonzalez+etal2016}; such massive discs are common around young
Class 0 and Class I objects \citep{Tychoniec+etal2018}. However, surveys of more
evolved protoplanetary discs with ages beyond approximately 1 Myr reveal typical
dust masses of between 1 and 100 $M_{\rm E}$ around solar-mass stars
\citep{Ansdell+etal2017}. Our nominal model with an accretion rate of
$10^{-8}\,M_\odot\,{\rm yr}^{-1}$ at 3 Myr contains 200 $M_{\rm E}$ of solids at
1 Myr and 30 $M_{\rm E}$ at 3 Myr, towards the high end of the survey results
presented in \cite{Ansdell+etal2017}.

The temporal decline of the gas accretion rate in the $\alpha$-disc model is set
by the viscous accretion time-scale over the initial disc size
\citep{Hartmann+etal1998}. Fixing the global viscosity coefficient at
$\alpha=0.01$, we find for our temperature profile that protoplanetary discs
starting with an accretion rate of $10^{-7}\,M_\odot\,{\rm yr}^{-1}$ and
reaching $5 \times 10^{-9}\,M_\odot\,{\rm yr}^{-1}$, $2 \times
10^{-9}\,M_\odot\,{\rm yr}^{-1}$ or $1 \times 10^{-9}\,M_\odot\,{\rm yr}^{-1}$
after 3 Myr have initial sizes of $R_1=51.3\,{\rm AU}, 24.3\,{\rm AU},15.0\,{\rm
AU}$, respectively. The initial masses of these discs are $0.074\,M_\odot,
0.039\,M_\odot, 0.026\,M_\odot$, respectively.

The resulting growth maps for a fixed pebble-to-gas flux ratio of $\xi=0.02$ are
shown in Figure \ref{f:growth_map_Mdot}. Decreasing the initial disc size and
mass leads to a significant decrease in the parameter space for forming
gas-giant planets. Hence both the initial disc size and the metallicity are
important factors in determining whether a protoplanetary disc will be able to
form gas-giant planets. \cite{Manara+etal2018} analysed the pebble masses in
protoplanetary discs around young stars in the Lupus region, with ages between 1
and 3 Myr, and concluded that these masses are too low to explain the statistics
of the observed exoplanet populations. They speculate that protoplanets must
grow even earlier in the life-time of the protoplanetary disc while there is
still plenty of pebbles to accrete. Our results agree with this picture:
planetary growth is very in efficient in the first Myr and will proceed even in
protoplanetary discs of initial sizes down to a few tens of AU.

\subsection{Varying the model parameters}
\label{s:variation}

In Figure \ref{f:growth_map_variation} we show the results of varying the
parameters of our model, to probe how sensitive planet formation is to the
physical properties of pebble growth, disc turbulence, and gap formation. All
growth maps were calculated for $\xi=0.02$, as this pebble-to-gas flux ratio
shows a large dependence on the physical parameters and thus makes differences
clear.

The top row of Figure \ref{f:growth_map_variation} shows the result of varying
the Stokes number. For larger Stokes number ${\rm St}=0.03$ the parameter space
for forming gas-giant planets increases relative to the nominal case with ${\rm
St}=0.01$. This increase in Stokes number affects the initial 3-D pebble
accretion stage most positively, since the accretion rate in the 3-D Hill regime
is proportional to the Stokes number. Lowering the Stokes number to ${\rm
St}=0.003$, on the other hand, has devastating effect on the ability of the
protoplaetary disc to form gas giants. The protoplanets are stuck at 3-D growth
below one Earth mass, unless they start very early.

In the middle row of Figure \ref{f:growth_map_variation} we increase the
turbulent viscosity (and the turbulent diffusion coefficient $\delta$) to
$\alpha_{\rm v}=3 \times 10^{-4}$ and $\alpha_{\rm v}=10^{-3}$.
\cite{Pinte+etal2016} inferred $\delta \sim 3 \times 10^{-4}$ from measurements
of the scale-height of the pebble layer. The translation from particle layer
scale height to turbulent diffusion coefficient nevertheless depends on the
assumed Stokes number. The high value of $\alpha_{\rm v}$ (and hence of
$\delta$) push planet formation to earlier times than for our nominal case of
$\alpha_{\rm v}=10^{-4}$ used in Figure \ref{f:growth_maps}.

Finally, we show in the bottom row of Figure \ref{f:growth_map_variation} two
additional experiments with gap formation. In the left panel we use the 50\% gap
mass from \cite{Kanagawa+etal2018} and the pebble isolation mass from the 3-D
simulations of \cite{Bitsch+etal2018b}. Taking these measurements at face value,
it is clear from Figure \ref{f:PIM_GTM_Mp} that the pebble isolation mass is
actually higher than the 50\% gap mass at $\alpha_{\rm v}=10^{-4}$! Since the
50\% gap mass is only determined by the relative height of the gap, while the
pebble isolation mass is set by both the gap height {\it and} shape, it could be
that gap edge instabilities would broaden the gap in such a way that the pebble
isolation mass comes after the 50\% gap mass. The result is a major reduction in
planetary migration. The final planetary mass is also reduced; this is a
consequence of the reduced migration, since the gas accretion rates are higher
in the inner regions of the protoplanetary disc where $H/r$ is low, according to
our adopted gas accretion model from \cite{TanigawaTanaka2016}. In the right
panel we assume instead that the 50\% gap mass is given by
\cite{Kanagawa+etal2018}, while the pebble isolation mass is 2.3 times lower.
Hence the pebble isolation mass is now only of the order of one Earth mass,
resulting in both reduced gas accretion rates (due to the small core mass) and
reduced migration (due to the small gap mass).

\section{Summary and discussion}

In this paper we have addressed the question of how planetary growth is able to
outperform migration. Type I migration likely operates at its full strength in
the outer regions of the protoplanetary disc where planetary cores grow to form
gas giants and ice giants, in contrast to the viscously heated inner regions of
the protoplanetary disc where the positive corotation torques slow down or
reverse the migration of super-Earth cores
\citep{BitschJohansen2016,Brasser+etal2017}.

We derived analytical growth tracks for a protoplanet undergoing pebble
accretion while migrating towards the star, under the assumptions that the
pebble Stokes number is a constant both in time and in space, that the pebble
mid-plane layer is sufficiently thin for pebble accretion to take place in 2-D,
and that the mass flux of pebbles is a constant ratio of the radial gas flow.
We used these analytical expressions to derive the radial location where the
growing core reaches the pebble isolation mass.  Our analytical growth tracks
demonstrate that protoplanets undergo substantial migration during their growth
towards the pebble isolation mass and that the location of reaching the pebble
isolation mass increases as the square root of the starting location.

An important new ingredient in our model, compared to previous studies of pebble
accretion and migration \citep{Bitsch+etal2015b,Matsumura+etal2017}, is the
reduction of the migration torque due to the growing planetary gap
\citep{Kanagawa+etal2018}. We identify here the close connection between the
50\% gap mass, $M_{\rm gap}$, of \cite{Kanagawa+etal2018} and the pebble
isolation mass, $M_{\rm iso}$, of \cite{Lambrechts+etal2014}. From numerical
simulations of low-mass planets embedded in a 1-D model of a protoplanetary
disc, we infer that $M_{\rm gap} \approx 2.3 M_{\rm iso}$. Reaching first the
pebble isolation mass and slightly later the gap transition mass thus signifies
three important events in the growth of a protoplanet: (a) the end of the
accretion of pebbles, (b) the beginning of gas contraction and (c) the
transition to a slow-down in the migration caused by the deepening gap.  Gas
accretion can therefore take place over just a few astronomical units of
migration. We emphasize that the connection between the 50\% gap mass and the
pebble isolation mass is derived here from 1-D simulations and that dedicated
2-D and 3-D simulations will be needed in the future to measure whether this
simplified relationship holds. At low viscosity ($\alpha < 10^{-3}$) our
measured pebble isolation mass is significantly lower than the measurements from
3-D simulations \citep{Bitsch+etal2018b}; this may be either due to gap edge
instabilities that are not captured in the 1-D approach
\citep{HallamPaardekooper2017} or due to the limited simulation time that can be
afforded in 3-D.

We have demonstrated that protoplanets can grow to gas-giant planets in models
where pebbles are relatively small, with a Stokes number of 0.01 corresponding
to approximately millimeter-sized particles in the inner regions of the
protoplanetary disc, and the pebble-to-gas flux ratio is in the nominal range
$\xi=0.01-0.02$, in contrast to the models of \cite{Bitsch+etal2015b},
\cite{Bitsch+etal2018a} and \cite{Brugger+etal2018} where pebbles were allowed
to grow to the radial drift barrier. The radial drift of the small pebbles
considered in this work is dominated by the advection with the gas accretion
flow -- and the gas and pebble components of the protoplanetary disc therefore
drain onto the central star on an approximately similar time-scale. This way we
avoid the radial drift problem where large pebbles drift out of the
protoplanetary disc on a time-scale that is much shorter than the gas accretion
time-scale \citep{Brauer+etal2007}.

Our results show that the cores of cold gas giants ending in 5--10 AU orbits
must generally start their assembly from planetesimals forming in the 15--30 AU
region. This raises the question of why planetesimals forming at such distances
would be the seeds of the gas giants.  The water ice line has been demonstrated
to be a preferred location for the formation of planetesimals
\citep{RosJohansen2013,IdaGuillot2016,SchoonenbergOrmel2017,DrazkowskaAlibert2017}.
If the ice lines of more volatile species, e.g.\ CO \citep{Qi+etal2013}, are
equally prone to forming planetesimals, then this may explain why the cores of
the gas-giants in the Solar System started as seeds far away from the Sun and
subsequently migrated to their current orbits.

\begin{acknowledgements}

The authors thank the anonymous referee whose report helped us to improve the
original manuscript. We would furthermore like to thank Hidekazu Tanaka,
Kazuhiro Kanagawa, Bertram Bitsch and Christoph Mordasini for inspiring
discussions. AJ is grateful to the Earth-Life Science Institute (ELSI) at Tokyo
Institute of Technology for hosting his research visit during March 2018. AJ
further thanks the Swedish Research Council (grant 2014-5775), the Knut and
Alice Wallenberg Foundation (grants 2012.0150, 2014.0017) and the European
Research Council (ERC Consolidator Grant 724687-PLANETESYS) for research
support. SI is supported by JSPS grant 15H02065 and MEXT grant 18H05438.

\end{acknowledgements}

\appendix

\section{Life-time of the pebble component of protoplanetary discs}
\label{s:pebble_disc}

Protoplanetary discs are efficient at converting dust to pebbles, due to their
high gas densities and weak turbulence. \cite{Testi+etal2003} and
\cite{Wilner+etal2005} inferred the presence of cm-sized pebbles in
protoplanetary discs around CQ Tau and TW Hya, respectively, from the spectral
energy distribution. Many other protoplanetary discs have been resolved at cm
wavelengths, including AS 209 \citep{Perez+etal2012}, CQ Tau
\citep{Trotta+etal2013} and HL Tau \citep{Carrasco-Gonzalez+etal2016}, to reveal
mm-sized pebbles in the outer parts of the discs and cm-sized pebbles in the
inner regions. \cite{Ansdell+etal2017} presented extensive survey results on
dust masses of protoplanetary discs observed with ALMA and found generally
1-100 Earth masses of mm-sized pebbles remaining in relatively evolved discs
around solar-type stars with ages in excess of one million years.

The presence of pebbles in protoplanetary is difficult to reconcile with the
short drift time-scales of pebbles \citep{Brauer+etal2007}.
\cite{LambrechtsJohansen2014} proposed that the delayed growth from dust to
pebbles in the low-density environment of the protoplanetary disc beyond 100 AU
provides a mechanism to maintain a long life-time of the pebble disc. The mass
flux of pebbles is determined not by the radial drift of the large pebbles close
to the star, but by the time-scale to grow to drifting sizes in the outer disc.
However, their model did not incorporate the exponentially tapered outer regions
of the protoplanetary disc, which form as a natural consequence of outwards
angular momentum transport \citep{Pringle1981}.
\begin{figure}
  \begin{center}
    \includegraphics[type=pdf,ext=.pdf,read=.pdf,width=0.9\linewidth]{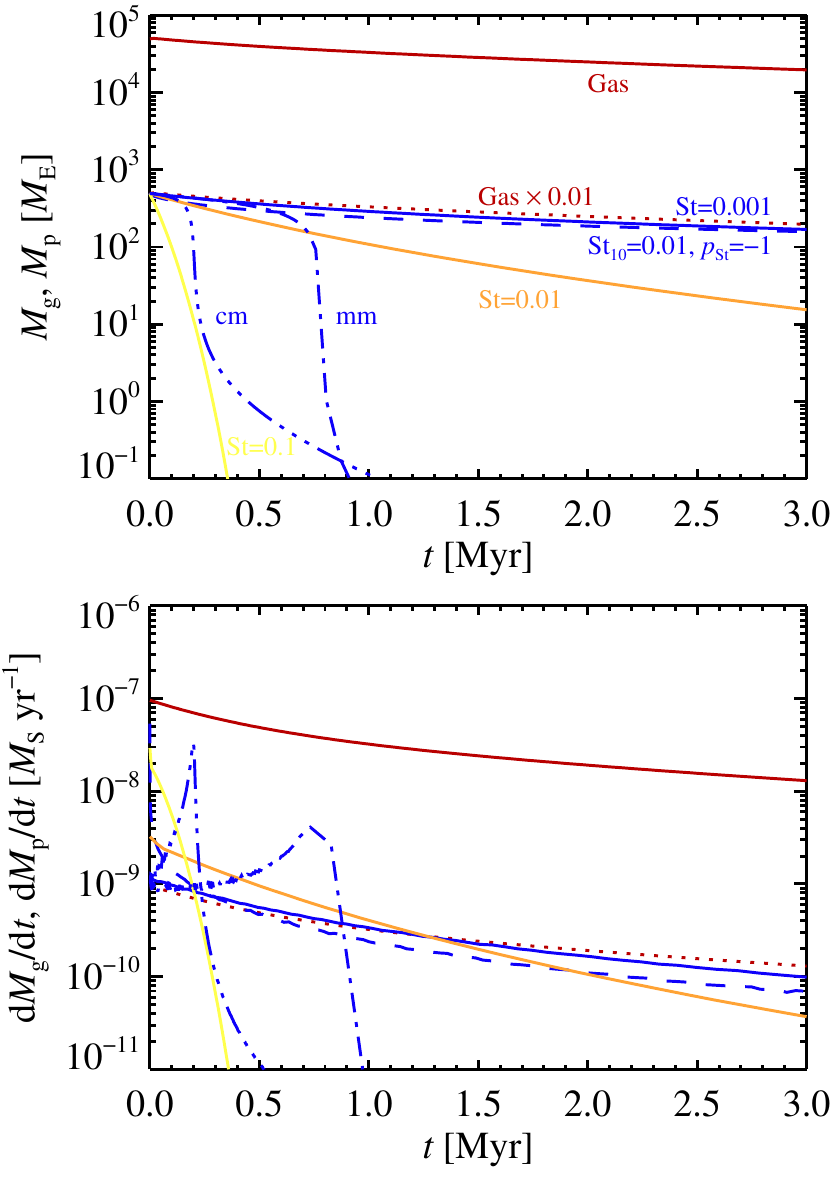}
  \end{center}
  \caption{The gas and pebble masses as a function of time (top panel) and the
  mass accretion rates of gas and pebbles (bottom panel). The different pebble
  models are explained in the main text. Protoplanetary discs with large pebbles
  -- of mm or cm sizes or with constant Stokes number of 0.1 -- drain their
  pebbles out of the disc in much less than one million years. The mass
  accretion rates of such large pebbles is very high during this period. The
  models with a lower Stokes number are better able to maintain their pebbles
  for the entire life-time of the protoplanetary disc.}
  \label{f:Mgas_Mdust_t_R90.8}
\end{figure}
\begin{figure}
  \begin{center}
    \includegraphics[type=pdf,ext=.pdf,read=.pdf,width=0.9\linewidth]{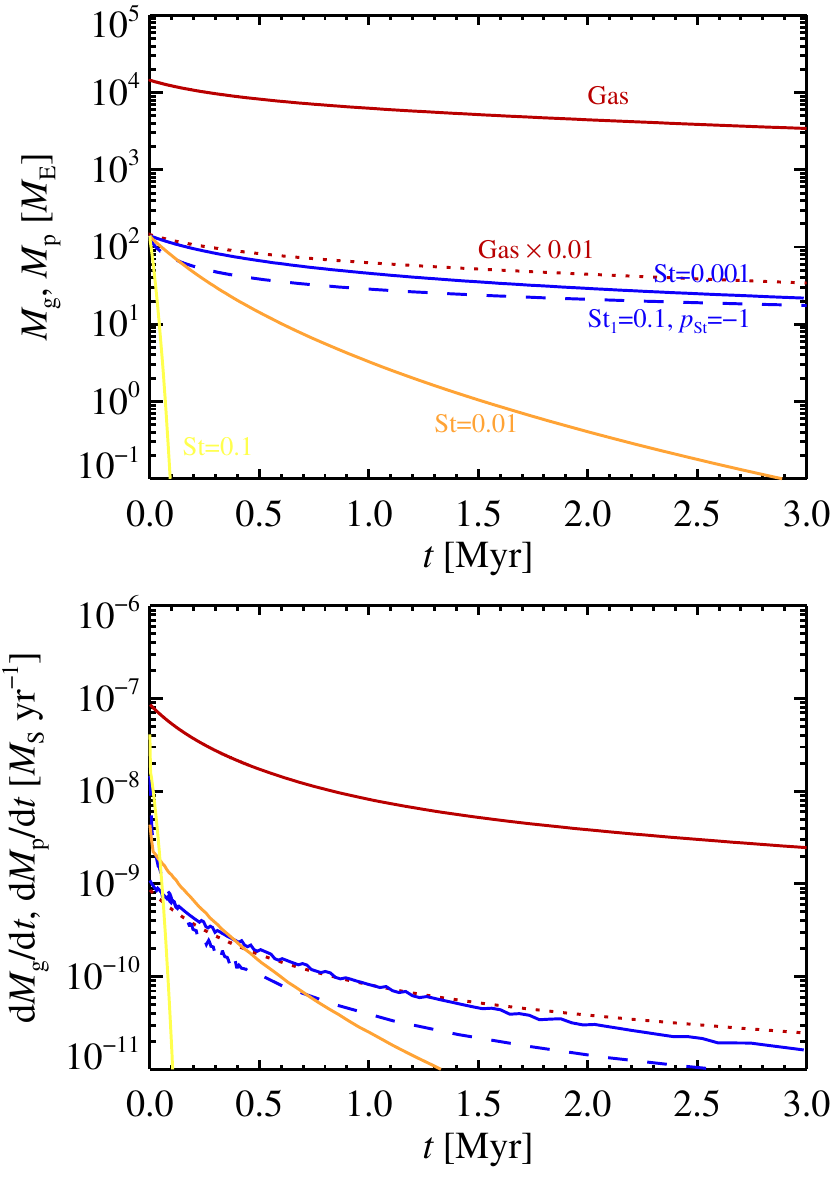}
  \end{center}
  \caption{The gas and pebble masses (top panel) and accretion rates (bottom
  panel) for a disc of small initial size $R_1=24.3\,{\rm AU}$ and initial mass
  $M=0.039\,M_\odot$. Pebbles of ${\rm St}=0.01$ are drained from the disc
  within 1 Myr due to the small initial disc size. However, smaller pebbles and
  pebbles with a radially declining Stokes number follow the temporal decay
  profile of the gas and are hence depleted on the same time-scale as the gas.}
  \label{f:Mgas_Mdust_t_R24.3}
\end{figure}

We explore here the evolution of drifting pebbles in an $\alpha$-model of a
protoplanetary disc. We let the gas evolve according to the analytical viscous
disc equations given in \cite{Hartmann+etal1998} and simulate the motion of
20,000 pebble superparticles, with the mass of each pebble swarm set to give an
initial dust-to-gas ratio $Z=0.01$ everywhere.  In Figure
\ref{f:Mgas_Mdust_t_R90.8} we show the total gas and pebble masses (top panel)
and mass accretion rates (bottom panel) for a disc that evolves from
$\dot{M}=10^{-7}\,M_\odot\,{\rm yr^{-1}}$ to $\dot{M}=10^{-8}\,M_\odot\,{\rm
yr^{-1}}$ over 3 Myr. We adopt $\alpha=10^{-2}$ for the disc evolution, so that
the disc has an initial size of $R_1=90.8$ AU. We use six models for the pebble
sizes: (i) constant Stokes number ${\rm St}=0.001$, (ii) constant Stokes number
${\rm St}=0.01$, (iii) constant Stokes number ${\rm St}=0.1$, (iv) constant
pebble size $R=1\,{\rm mm}$, (v) constant pebble size $R=1\,{\rm cm}$ and (vi)
variable Stokes number ${\rm St}=0.01 [r/(10\,{\rm AU})]^{-1}$.  The gas mass in
these $\alpha$-disc models falls only slowly with time; this is due to the
power-law nature of the temporal evolution of the accretion rate.
Protoplanetary discs may need FUV photoevaporation to dissipate these viscously
spreading outer regions \citep{Gorti+etal2015}.

The simulations presented in Figure \ref{f:Mgas_Mdust_t_R90.8} with mm-sized
pebbles, cm-sized pebbles or pebbles with ${\rm St}=0.1$ are quickly drained out
of the protoplanetary disc, in much less than one million years. The pebble
fluxes are initially very high, with $\xi=\dot{M}_{\rm p}/\dot{M}_{\rm g}$
reaching values above 0.1. While such mass fluxes could certainly lead to very
high pebble accretion rates, the rapid depletion of pebbles appears in conflict
with observations of $1$--$100$ $M_{\rm E}$ of pebbles in evolved protoplanetary
discs around solar-mass stars \citep{Ansdell+etal2017}.

The models with ${\rm St}=0.01$, ${\rm St}=0.001$ or power-law Stokes number are
more successful at maintaining the pebbles for the life-time of the
protoplanetary disc. Here the low drift rates of the small pebbles, as well as
the outwards drift of pebbles in the viscously expanding outer regions of the
protoplanetary disc, keep the pebble flux at a level of around 1\% of the gas
flux. The model with ${\rm St}=0.01$ at 10 AU and falling inversely
proportionally to the distance could have its physical foundation in a turbulent
$\alpha$ that increases in the outer disc, limiting the particle growth there
\citep[as proposed in][]{Ida+etal2016}. For this model we also limited the
smallest particle size to one micron.

In Figure \ref{f:Mgas_Mdust_t_R24.3} we show the gas and pebble evolution for an
initial disc size of $24.3\,{\rm AU}$. The initial disc mass is thus only
$0.039\,M_\odot$, towards the low end for the distribution of masses of embedded
Class 0 and Class I objects \citep{Tychoniec+etal2018}. Even small pebbles with
${\rm St}=0.01$ only survive for less than 1 Myr in such a small disc. Tiny
pebbles with ${\rm St}=0.001$ nevertheless follow the gas accretion profile
closely, as does the model with a Stokes number that decreases with distance.

While the realism of the $\alpha$-model that we use here for the gas is
debatable in light of modern disc-wind models for accretion
\citep{BaiStone2013,Bethune+etal2017}, overall we find that the presence of
pebbles in protoplanetary discs for million years may be determined by the
limited growth of particles in the outer regions of the protoplanetary disc that
acts as a drift bottleneck. Such small and cold dust far from star would be hard
to detect but may reveal its presence in scattered light images of
protoplanetary discs that often extends to sizes much larger than at mm
wavelengths \citep[e.g.][]{vanBoekel+etal2017}.

\section{Analytical expression for the growth track time}
\label{s:growth_time}

In this Appendix we derive the time associated with the analytical growth track
derived in Section \ref{s:core_growth}. We replace the pebble surface density
$\varSigma_{\rm p}$ in equation (\ref{eq:Mdot}) by equation (\ref{eq:Mdotp}) and
split out the specific dependence on $M$ and $r$ to obtain
\begin{eqnarray}
  \dot{M} &=& 2 \left( \frac{\rm St}{0.1} \right)^{2/3} G M_\star
  (3 M_\star)^{-2/3} \nonumber \\
  && \times \frac{\xi \dot{M}_{\rm g}(t)}{2 \pi [\chi {\rm St}+(3/2)\alpha]
  c_{\rm s1}^2 {\rm AU}^{\zeta}} M^{2/3} r^{\zeta-1} \, .
\end{eqnarray}
We need now a model for the evolution of the protoplanetary disc. We use the
standard $\alpha$-disc evolution model of Section \ref{s:full_growth}. The
turbulent viscosity follows the power-law
\begin{equation}
  \nu = \alpha c_{\rm s} H \propto r^{\gamma} \, .
\end{equation}
Here the power-law index $\gamma$ can be written as
\begin{equation}
  \gamma = (3/2) - \zeta \, .
\end{equation}
The gas accretion onto the star evolves as \citep{Hartmann+etal1998}
\begin{equation}
  \dot{M}_{\rm g} = \dot{M}_{\rm g0} T^{-(5/2-\gamma)/(2-\gamma)} \,
  ,
\end{equation}
where the non-dimensional time is
\begin{equation}
  T = t/t_{\rm s} + 1 \, ,
  \label{eq:AT}
\end{equation}
the characteristic time is
\begin{equation}
  t_{\rm s} = \frac{1}{3(2-\gamma)^2} \frac{R_1^2}{\nu_1} \, ,
\end{equation}
and $R_1$ is the initial disc size and $\nu_1$ the viscosity at that location.
We now insert the inverse solution for the growth track, $r(M)$, and separate
the variables $M$ and $r$ to yield the equation
\begin{eqnarray}
  r_0^{1-\zeta} M^{-2/3} \left( 1-\frac{M^{4/3}-M_0^{4/3}}{M_{\rm
  max}^{4/3}-M_0^{4/3}} \right) {\rm d}M = \nonumber \\ \frac{2 ({\rm
  St}/0.1)^{2/3} G M_\star (3 M_\star)^{-2/3} \xi \dot{M}_{\rm g0}}{2 \pi[\chi
  {\rm St} +(3/2) \alpha] c_{\rm s1}^2 {\rm AU}^{\zeta }}
  T^{-(5/2-\gamma)/(2-\gamma)} t_{\rm s} {\rm d} T \, .
\end{eqnarray}
We integrate both sides to yield
\begin{eqnarray}
  r_0^{1-\zeta} \left[ \frac{-(3/5) (M^{5/3}
  - M_0^{5/3}) + 3 (M^{1/3} - M_0^{1/3})M_{\rm max}^{4/3}}{M_{\rm max}^{4/3} -
    M_0^{4/3}} \right] = \nonumber \\ \frac{2 ({\rm St}/0.1)^{2/3} G M_\star (3
    M_\star)^{-2/3} \xi \dot{M}_{\rm g0}}{2 \pi[\chi {\rm St} +(3/2) \alpha]
    c_{\rm s1}^2 {\rm AU}^{\zeta}} \times \nonumber \\
    \frac{[T^{-(5/2-\gamma)/(2-\gamma)+1} -
    T_0^{-(5/2-\gamma)/(2-\gamma)+1} ]t_{\rm s}}{-(5/2-\gamma)/(2-\gamma)+1} \,
    .
    \label{eq:time}
\end{eqnarray}
We now divide the equation by the solution at $r=0$, $M(T_{\rm max})=M_{\rm
max}$, to give the simpler expression
\begin{eqnarray}
  \frac{-(3/5) (M^{5/3}
  - M_0^{5/3}) + 3 (M^{1/3} - M_0^{1/3})M_{\rm max}^{4/3}}{-(3/5) (M_{\rm
    max}^{5/3}
  - M_0^{5/3}) + 3 (M_{\rm max}^{1/3} - M_0^{1/3})M_{\rm max}^{4/3}} = \nonumber
    \\ \frac{T^{-(5/2-\gamma)/(2-\gamma)+1} -
    T_0^{-(5/2-\gamma)/(2-\gamma)+1}}{T_{\rm max}^{-(5/2-\gamma)/(2-\gamma)+1} -
    T_0^{-(5/2-\gamma)/(2-\gamma)+1} } \, .
  \label{eq:time2}
\end{eqnarray}
Note that $T_{\rm max}$ may be complex if the planet never reaches $r=0$.  The
solution can easily be solved for $T$ for a given $M$, which can be converted to
$t$ using the viscous disc expression from equation (\ref{eq:AT}). The
analytical expression for $T_{\rm max}$ is obtained by inserting $M=M_{\rm max}$
in equation (\ref{eq:time}).

If we assume for simplicity that $\dot{M}_{\rm g}$ is constant, then we can
replace the complicated function of $T$ appearing on the right-hand-side of
equation (\ref{eq:time}) by $t-t_0$. In that case (setting $M_0=t_0=0$ also) we
have for $t_{\rm max}$ the simple expression
\begin{eqnarray}
  t_{\rm max} &=& r_0^{1-\zeta} (12/5) M_{\rm max}^{1/3} \nonumber \\ && \times
  \frac{2 \pi (3/2) [(2/3) ({\rm St}/\alpha) \chi + 1] \alpha c_{\rm s1}^2 {\rm
  AU}^{\zeta}}{2 ({\rm St}/0.1)^{2/3} G M_\star (3 M_\star)^{-2/3} \xi
  \dot{M}_{\rm g0}} \, .
\end{eqnarray}
The time-evolution of the growth track follows in the limit $M_0=t_0=0$
as $t/t_{\rm max}=(5/3) (M/M_{\rm max})^{1/3}-(1/4) (M/M_{\rm max})^{5/3}$. We
can scale the time-scale to migrate to the star, $t_{\rm max}$, to typical disc
values,
\begin{eqnarray}
  t_{\rm max} &=& 0.29\,{\rm Myr}\,  \left\{ \frac{[(2/3) ({\rm St}/\alpha) \chi
  + 1]/2.9}{({\rm St}/0.01)^{2/3}} \right\}  \left( \frac{\xi}{0.01}
  \right)^{-1} \nonumber \\
  && \times \left( \frac{M_\star}{M_\odot} \right)^{-1/3} \left( \frac{c_{\rm
  s1}}{6.5 \times 10^2\,{\rm m\,s^{-1}}} \right)^2
  \nonumber \\
  && \times \left( \frac{\alpha}{0.01} \right) \left( \frac{\dot{M}_{\rm
  g0}}{10^{-7}\,M_\odot\,{\rm yr^{-1}}} \right)^{-1} \nonumber \\
  && \times \left( \frac{M_{\rm max}}{10\,M_{\rm E}} \right)^{1/3} \left(
  \frac{r_0}{25\,{\rm AU}} \right)^{1-\zeta} \, .
  \label{eq:tmaxs}
\end{eqnarray}
The relevant value of $M_{\rm max}$ for the given parameters can be inserted
from equation (\ref{eq:Mmaxs}). However, we note that $t_{\rm max}$ depends
weakly on the maximum mass, as $M_{\rm max}^{1/3}$. Hence the parameter
dependencies in equation (\ref{eq:Mmaxs}) are much weaker than those explicit in
equation (\ref{eq:tmaxs}) above. One can also approximately use $t_{\rm max}$ as
the time to reach the isolation mass, as the isolation mass and the maximum mass
are both reached after significant migration when $r_0 \gg 10\,{\rm AU}$. Under
all circumstances does $t_{\rm max}$ give an upper limit to the time of reaching
the isolation mass.

\end{document}